\documentclass[prd,superscriptaddress,floatfix,notitlepage,nofootinbib,reprint]{revtex4-1}  

\usepackage{graphicx}  
\usepackage{dcolumn}   
\usepackage{bm}        
\usepackage{amsmath,amssymb}   
\usepackage{aas_macros}
\usepackage{multirow}
\usepackage{color}
\usepackage{verbatim}
\usepackage{url}
\usepackage{hyperref}

\newcommand{\mr}{\mathrm}

\newcommand{\bea}{\begin{eqnarray}}
\newcommand{\eea}{\end{eqnarray}}
\newcommand{\bmp}{\bm{\psi}}
\newcommand{\bmv}{\bm{v}}
\newcommand{\bmk}{\bm{k}}
\newcommand{\bmx}{\bm{x}}
\newcommand{\bms}{\bm{s}}
\newcommand{\bmq}{\bm{q}}
\newcommand{\bmxi}{\bm{\xi}}
\newcommand{\la}{\langle}
\newcommand{\ra}{\rangle}

\begin{document}
\widetext

\title{Nonlinear reconstruction of redshift space distortions}

\author{Hong-Ming Zhu}
    \email[]{hmzhu@berkeley.edu}
\affiliation{Berkeley Center for Cosmological Physics and Department of Physics,
University of California, Berkeley, California 94720, USA}
\affiliation{Key Laboratory for Computational Astrophysics, National Astronomical Observatories, Chinese Academy of Sciences, 20A Datun Road, Beijing 100012, China}
\affiliation{University of Chinese Academy of Sciences, Beijing 100049, China}

\author{Yu Yu}
\affiliation{Department of Astronomy, Shanghai Jiao Tong University, 800 Dongchuan Road, Shanghai 200240, China}
\affiliation{Key Laboratory for Research in Galaxies and Cosmology,
Shanghai Astronomical Observatory, Chinese Academy of Sciences,
80 Nandan Road, Shanghai 200030, China}

\author{Ue-Li Pen}
\affiliation{Canadian Institute for Theoretical Astrophysics, University of Toronto, 60 St. George Street, Toronto, Ontario M5S 3H8, Canada}
\affiliation{Dunlap Institute for Astronomy and Astrophysics, University of Toronto, 50 St. George Street, Toronto, Ontario M5S 3H4, Canada}
\affiliation{Canadian Institute for Advanced Research, CIFAR Program in Gravitation and Cosmology, Toronto, Ontario M5G 1Z8, Canada}
\affiliation{Perimeter Institute for Theoretical Physics, 31 Caroline Street North, Waterloo, Ontario, N2L 2Y5, Canada}

\date{\today}

\begin{abstract}
We apply nonlinear reconstruction to the dark matter density field in redshift 
space and solve for the nonlinear mapping from the initial Lagrangian position
to the final redshift space position.
The reconstructed anisotropic field inferred from the nonlinear 
displacement correlates with the linear initial conditions to much smaller
scales than the redshift space density field.
The number of linear modes in the density field is improved by a factor of 
$30$--$40$ after reconstruction. 
We thus expect this reconstruction approach to substantially expand 
the cosmological information including baryon acoustic oscillations and redshift
space distortions for dense low-redshift 
large scale structure surveys including for example SDSS main sample, DESI 
BGS, and 21 cm intensity mapping surveys.
\end{abstract}

\pacs{}
\maketitle

\section{Introduction}
Measuring the three-dimensional large-scale structure of the Universe provides
a powerful method to probe particle physics and cosmology.
Precision measurements of baryon acoustic oscillations (BAO) and redshift space
distortions (RSD) can constrain dark energy models and modified theories of
gravity \cite{2016sdss,2017F,2017R,2016V,2017F2,2017G,2017S,2016S,2017Z}. 
The ongoing and upcoming surveys are mapping large swaths of the visible 
Universe \cite{2016sdss,CHIME,tianlai,DESI,pfs}.
However, the precision of the measured cosmological parameters (e.g. the BAO 
scale and the growth rate) is usually limited by the strong non-Gaussianities
of the nonlinear data, which prevent a simple mapping to the linear initial 
conditions which are predicted by cosmological theories. 
Therefore, to better extract cosmological information from an observed nonlinear
map, it is crucial to understand the nonlinearities that make the observed 
signatures deviate from the theoretical predictions and how to reduce these 
nonlinearities in the nonlinear map to obtain better statistics.

Reconstruction methods have been developed to reverse the nonlinear degradation
of the BAO signal \cite{2007bao,2012TZ2,2015marcel,2017BAOP}. 
The current standard BAO reconstruction method uses a linear mapping to reduce
the shift nonlinearities in the observed nonlinear field \cite{2007bao}.
The linear method is successfully applied to galaxy surveys and improves the
measurements of the BAO scale \cite{2012nikhil,2012Anderson,2014K,2014Anderson,2014T,2015Ross,2017F,2017R}.
However, the density field reconstruction methods based on the linearized 
continuity equation only capture the shift terms from large-scale
linear bulk flows \cite{2007ESW,2009PWC,2009NWP,2012TZ,2014Tassev,2016Seo}.
Recently a new reconstruction method has been proposed in Ref. \cite{2016HMZ}. 
(See also Refs. \cite{2017Marcel,2017Shi} for different implementations  
based on the same principle.)
The new method solves a nonlinear bijective mapping between the Eulerian 
coordinate system and a new coordinate system, where the mass per volume 
element is constant \cite{2016HMZ}. 
The nonlinear mapping provides a significantly better estimate of the nonlinear
displacement compared to standard reconstruction.
The linear density field inferred from the nonlinear bijective mapping gives a
much better correlation with the linear initial conditions \cite{2016HMZ}.
These nonlinear methods are expected to substantially improve BAO measurements 
in the near future.

Most nonlinear reconstruction methods developed recently focus on improving 
measurements of the linear BAO signal \cite{2016HMZ,2017Marcel,2017Shi,2017US}.
The current standard reconstruction also only works for reversing the nonlinear
degradation of the BAO feature in galaxy redshift surveys, though different
conventions for the RSD effect are adopted. (See Ref. \cite{2016Seo} for more
discussions about different RSD conventions.) 
The RSD effect is also one of the most powerful probes of the Universe, 
complementary to the BAO measurements \cite{1972MNRAS.156P...1J,1977ApJ...212L...3S,1980lssu.book.....P,1987MNRAS.227....1K,1994MNRAS.267.1020P,1996MNRAS.282..877B}. 
Precise measurements of the growth rate
of structure from the RSD effect can provide tests of general relativity. 
While the distinct BAO feature can be measured easily, extracting information
from RSD measurements requires the modeling of the power spectrum amplitude.
Thus the analysis of RSD is often limited to much smaller wave numbers
(e.g. $k<0.15\ h/\mr{Mpc}$ compared to $k<0.3\ h/\mr{Mpc}$ for the BAO-only
analysis \cite{2017F,2017F2}).
A lot of effort has been made to model the nonlinear RSD effect \cite{1995F,1998H,2001W,2001S,2002K,2004S,2006T,2007T,2010D,2010TNS,2011M,2011O,2011O2,2011S,2011J,2011GSM,2011DF,2012O,2012O2,2012K,2013Z,2013Z2,2014I,2015W,2015B,2016B,2016J,2017H}.
However, even if the nonlinear RSD effect can be modeled to nonlinear scales,
this still does not improve the power spectrum information content of a 
nonlinear map.
Present RSD analysis is limited by the strong nonlinearities of the density 
and velocity fields and the nonlinear mapping from real to redshift space, with
a small amount of cosmological information being extracted from galaxy surveys.
Another possible way to handle the nonlinear RSD effect would be trying to 
reduce non-Gaussianities of the nonlinear density field in redshift space,
in order to unscramble the cosmological information encoded in higher order 
statistics of the nonlinear data and increase the power spectrum information 
content of the reconstructed linear map.

In this paper, we apply nonlinear reconstruction to the dark matter density 
field in redshift space and solve for the nonlinear mapping from the initial 
Lagrangian position to the final redshift space position.
The reconstructed anisotropic field inferred from the nonlinear 
displacement correlates with the linear initial conditions to much smaller
scales than the redshift space density field.
Therefore, we are able to significantly increase the constraining power of 
anisotropic galaxy clustering measurements by including small-scale density
fluctuations in the analysis and provide tighter constraints
on the structure growth rate.
The idea of RSD reconstruction proposed here would be crucial for improving
the RSD measurements from current and future galaxy surveys.

This paper is organized as follows. In Sec. \ref{sec:mot}, we describe the 
motivation for reconstruction of RSD.
Section \ref{sec:prf} describes the performance tests of reconstruction and 
shows the reconstruction results.
In Sec. \ref{sec:app}, we describe some applications of reconstruction.
We conclude in Sec. \ref{sec:cls}.
Appendix \ref{appendix:A} shows the statistical properties of Lagrangian 
displacements. 
In Appendixes \ref{appendix:B} and \ref{appendix:C}, we present the 
study of Lagrangian velocities and shift velocities.
Appendix \ref{appendix:D} shows the RSD reconstruction results for halos.
Appendix \ref{appendix:E} describes how to implement nonlinear reconstruction
in the presence of mask.


\section{Physical motivation}
\label{sec:mot}

The Lagrangian space perturbation theories which capture the nonlinear shift 
terms provide a successful description of the matter distribution on large 
scales \cite{2008M,2008M2,LPT3,2014PSZ,2013CRW,2015VSB,2015SV,2015VWA,2016VCW,
2016matt,2016BSZ,2016BSZ2}.
In the Lagrangian picture of structure formation, the Eulerian position $\bmx$
of each mass element is given by the sum of its initial Lagrangian coordinate 
$\bmq$ and the subsequent displacement $\bmp(\bmq)$,
\bea
\bm{x}(\bm{q})=\bmq+\bm{\psi}(\bmq).
\eea
The displacement field which defines the mapping from initial Lagrangian 
coordinates to Eulerian coordinates is the key variable in Lagrangian space 
based perturbation theories.
Since the initial density field is almost uniform, the nonlinear density at the
Eulerian position $\bmx$ is 
\bea
\rho(\bmx)d^3x=\rho(\bmq)d^3q,
\eea
where $\rho(\bmq)\approx\mr{constant}$. The mass conservation equation predicts
the exact nonlinear density $\rho(\bmx)=d^3q/d^3x-1$ at $\bmx$ in the absence of
shell crossing. Therefore, the key ingredient of Lagrangian perturbation theory
calculation is to solve the nonlinear displacement field and calculate the 
Jacobian of the mapping from $\bmq$ to $\bmx$.

The redshift space position $\bms$ of an object is shifted from its real space 
position $\bmx$ by its peculiar velocity,
\bea
\bms(\bmx)=\bmx+\frac{v_z(\bmx)}{aH}\hat{z},
\eea
where $a$ is the scale factor and $H$ is the Hubble parameter. In this paper 
we adopt the plane-parallel approximation and take the line of sight direction 
to be the $z$-axis. The density field in redshift space can be obtained by 
imposing mass conservation,
\bea
\rho(\bms)d^3s=\rho(\bmx)d^3x.
\eea
Since the growth of displacement $\bmp(\bmq)$ is the Lagrangian velocity
$\bmv(\bmq)$, the Eulerian velocity $\bmv(\bmx)$ of a mass element at 
$\bmx=\bmq+\bmp(\bmq)$ is simply the Lagrangian velocity $\bmv(\bmq)$.
Thus we can define a mapping from the initial Lagrangian coordinates $\bmq$ to 
the redshift space coordinates $\bms$ as
\bea
\bms(\bmq)=\bmq+\bmp^s(\bmq),
\eea
where
\bea
\bmp^s(\bmq)=\bmp(\bmq)+{\bmv^s(\bmq)}/{aH}.
\eea
Here $\bmv^s(\bmq)\equiv[\hat{z}\cdot\bmv(\bmq)]\hat{z}$ is called the shift velocity.
The redshift space nonlinear density field is given by
\bea
\label{eq:mc_rsd}
\rho(\bms)d^3s=\rho(\bmq)d^3q.
\eea
Again the nonlinear density at $\bms$ predicted by Eq. (\ref{eq:mc_rsd})
with the redshift space displacement field $\bmp^s(\bmq)$ is exact before shell
crossing happens.

The nonlinear mapping from real to redshift space has been studied extensively,
which is necessary for modeling the nonlinear RSD effect. A successful nonlinear
RSD model can reduce theoretical systematics in the analysis. However, the 
cosmological information extracted from redshift space two-point statistics 
is still limited by the shift nonlinearities in the observed density field. 
In order to remove the nonlinear shift terms induced by the mapping from $\bmq$
to $\bmx$ and then from $\bmx$ to $\bms$, we can try to invert the nonlinear 
mapping and solve for the Lagrangian displacement and the shift velocity, which
are Lagrangian quantities that are not affected by shift terms.

The basic idea of nonlinear reconstruction is to build a bijective mapping between the redshift space coordinates $\bms$ and the potential isobaric coordinates $\bmxi$, where $\rho(\bmxi)d^3\xi$ is approximately constant.
We define a coordinate transformation that is a pure gradient,
\bea
s^i=\xi^{\mu}\delta^i_{\mu}+\frac{\partial\phi^s}{\partial\xi^{\nu}}\delta^{i\nu},
\eea
where $\phi^s(\bmxi)$ is the displacement potential. 
The potential isobaric gauge is unique as long as we require that the coordinate
transformation defined above is positive definite, i.e., $\mr{det}(\partial s^i/\partial\xi^{\alpha})>0$. 
It becomes analogous to synchronous gauge and Lagrangian coordinates before 
shell crossing, but allows a unique mapping even after shell crossing.
The unique displacement potential $\phi^s(\bmxi)$ consistent with the nonlinear
density and positive definite coordinate transformation can be solved using
the moving mesh approach, which evolves the coordinate system towards a state 
of constant mass per volume element, i.e., $\rho(\bmxi)d^3\xi=\mr{constant}$
\cite{1995ApJS..100..269P,1998ApJS..115...19P}.
See Ref. \cite{2016HMZ} for the details of this calculation.

The gradient of the reconstructed  displacement potential $\nabla_{\bmxi}\phi^s(\bmxi)$ provides an estimate of the redshift space displacement $\bmp^s(\bmq)$.
We define the negative Laplacian of the reconstructed displacement potential as
the reconstructed density field,
\bea
\label{eq:density}
\delta_r^s(\bmxi)=-\nabla_{\bmxi}\cdot\nabla_{\bmxi}\phi^s(\bmxi)=-\nabla^2_{\bmxi}\phi^s(\bmxi).
\eea
Note that the reconstructed density field is computed on the potential isobaric
coordinates instead of the redshift space coordinates.

Because of the limited degrees of freedom, we can only solve the scalar part 
of the nonlinear mapping, while the real nonlinear displacement also includes 
the curl part. 
The gradient and curl parts of the redshift space nonlinear displacement are
\bea
\bmp^s_E(\bmq)=\bmp_E(\bmq)+{\bmv^s_E(\bmq)}/{aH},
\eea
and
\bea
\bmp^s_B(\bmq)=\bmp_B(\bmq)+{\bmv^s_B(\bmq)}/{aH},
\eea
respectively, where 
\bea
\bmp_E(\bmk)=[\bmk\cdot\bmp(\bmk)]\bmk/k^2,\ \bmp_B(\bmk)=\bmp(\bmk)-\bmp_E(\bmk),
\eea
and 
\bea
\bmv^s_E(\bmk)=[\bmk\cdot\bmv^s(\bmk)]\bmk/k^2,\ \bmv_B^s(\bmk)=\bmv^s(\bmk)-\bmv^s_E(\bmk).
\eea
Since the displacement field is anisotropic in the presence of redshift space
distortions, the reconstructed displacement potential is also anisotropic.
In addition to the real space $E$-mode displacement $\bmp_E(\bmq)$, the solved
displacement also contains the $E$-mode shift velocity $\bmv_E^s(\bmq)/aH$.
The shift velocity is the most important observable in the RSD measurements 
since it describes the growth of the displacement, which provides cosmological 
information about the logarithmic growth rate $f=d\mr{ln}D/d\mr{ln}a$, where
$D$ is the linear growth factor.
This is new to all previous reconstruction algorithms since it is about  
reconstruction of the redshift space distortions.

Notice that the shift velocity $\bmv^s(\bmq)$ is different from $\bmv(\bmq)$, 
where the former corresponds to the $z$-component of the latter.
The Lagrangian velocities $\bmv_{E}$ and $\bmv_{B}$ describe the growth of the 
Lagrangian displacements $\bmp_{E}$ and $\bmp_{B}$, respectively. 
The statistical properties of the Lagrangian displacements have been studied in
Appendix \ref{appendix:A}, while the Lagrangian and shift velocities have been
investigated in Appendixes \ref{appendix:B} and \ref{appendix:C}.

\section{Performance tests}
\label{sec:prf}

To test the performance of reconstruction with redshift space density field, 
we use a set of ten $N$-body simulations run with the ${\tt CUBEP^3M}$ code 
\cite{2013code}. These simulations evolve $1024^3$ dark matter particles in 
a cubic box of length $600\ \mr{Mpc}/h$.
We take the snapshot at redshift $z=0$ and compute the density on a $512^3$ 
grid. We solve the displacement potential $\phi^s$ using the multigrid algorithm
and then compute the reconstructed density field using Eq. (\ref{eq:density}) 
in the potential isobaric gauge \cite{2016HMZ}.

To check convergence of the reconstruction results, we run a high-resolution
simulation of $2048^3$ particles with the same box size. 
We also perform reconstruction with the redshift space density field on a $512^3$ grid from the high-resolution simulation. We find that the reconstruction 
results with different simulation resolutions are almost indistinguishable. 
Therefore, we use the high-resolution simulation for visual comparison due to
its better image resolution and use the statistical averaged power spectra from
the other ten simulations to reduce statistical errors.

In the following subsections, we assess the performance of reconstruction in 
redshift space in terms of the density field slice, the correlation with the
initial conditions, the correlation with the nonlinear field, the shape of 
the power spectrum, and the correlation with the Lagrangian displacements and 
velocities.

\subsection{Nonlinear map with potential isobaric gauge}

\begin{figure}[tbp]
\begin{center}
\includegraphics[width=0.45\textwidth]{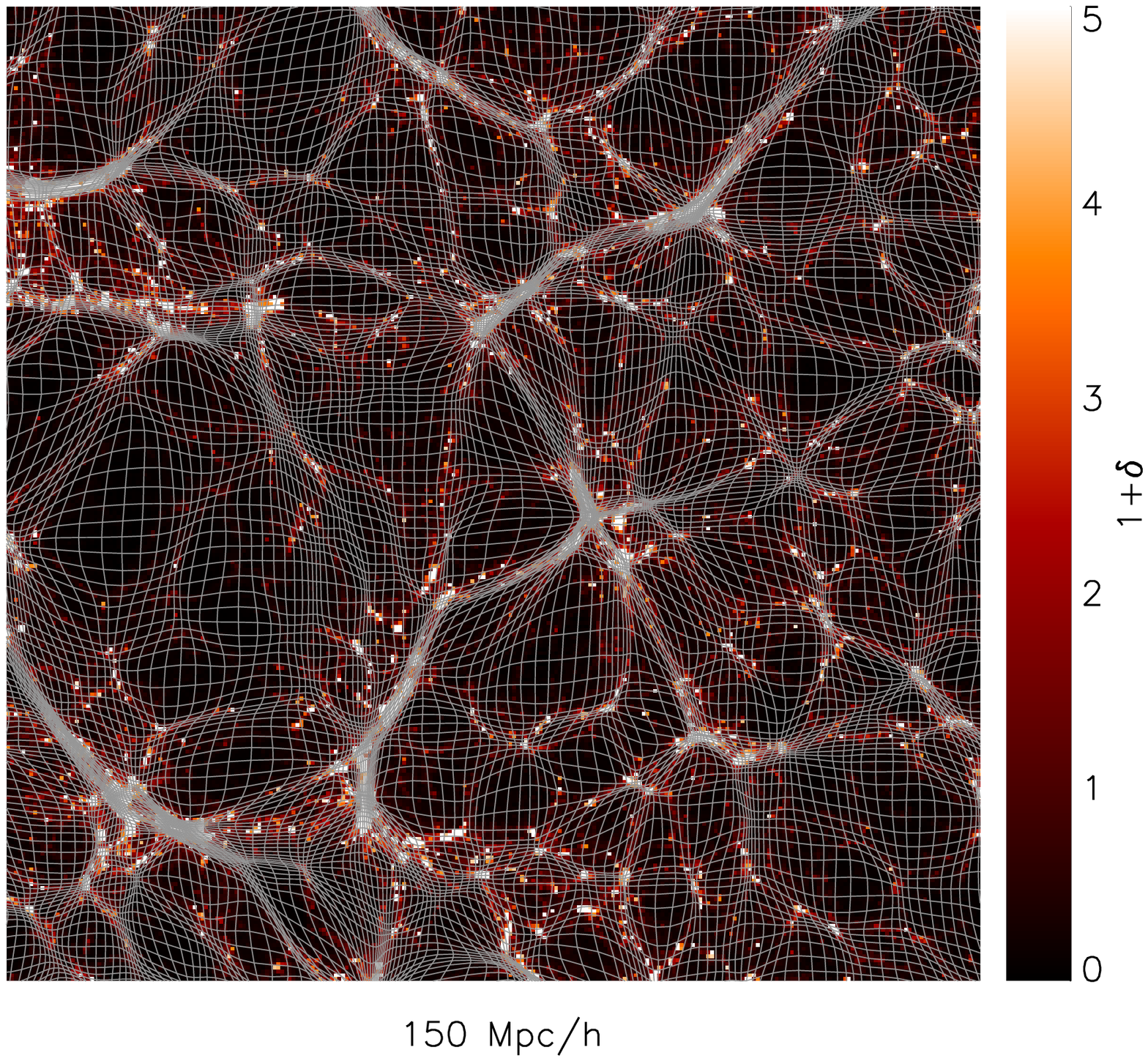}
\end{center}
\vspace{-0.7cm}
\caption{A slice of the nonlinear density field from the high-resolution 
    simulation without the RSD effect. The curvilinear grid shows the Eulerian
    coordinate of each grid point of the potential isobaric coordinate.}
\label{fig:map}
\end{figure}

\begin{figure}[tbp]
\begin{center}
\includegraphics[width=0.45\textwidth]{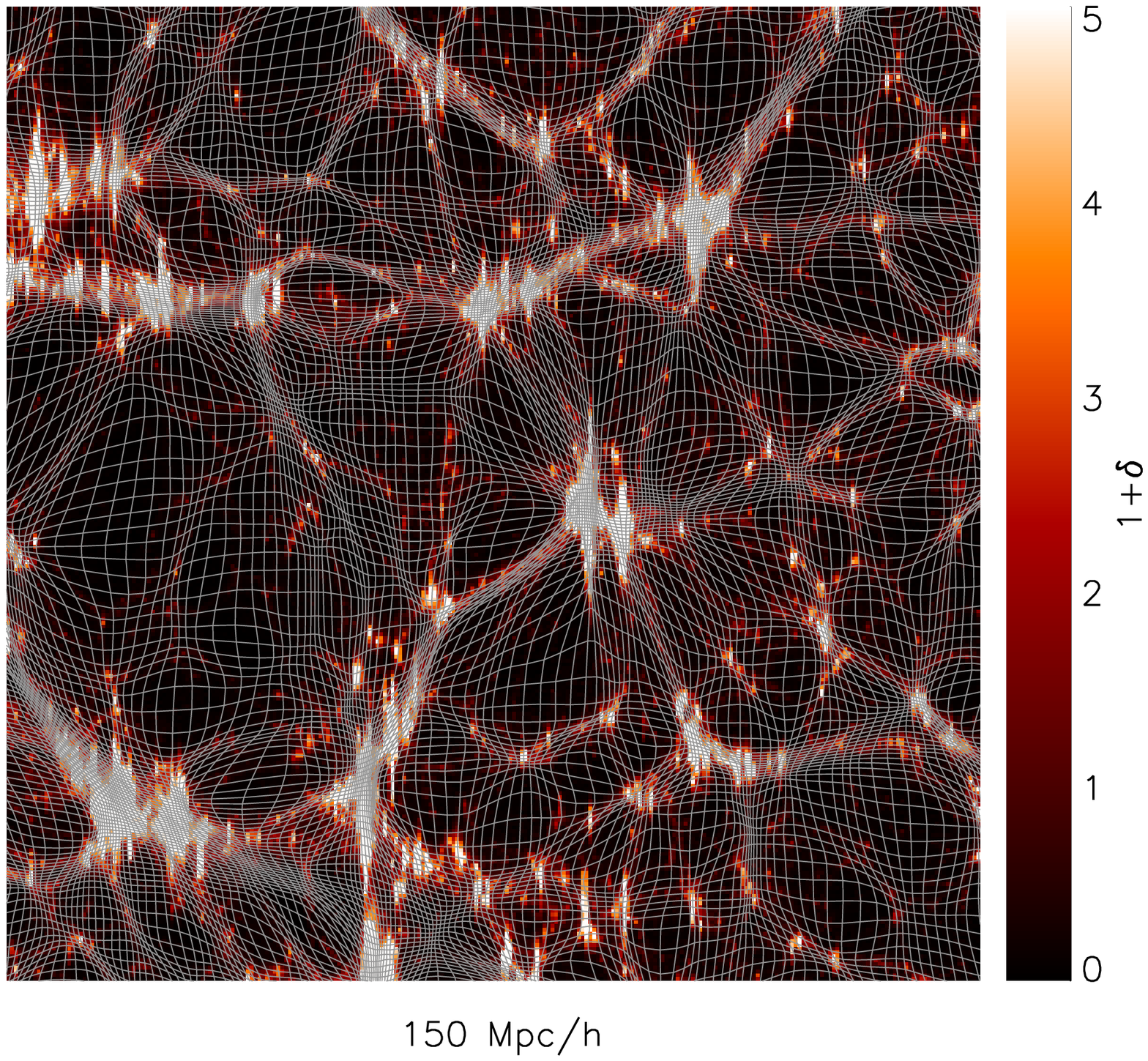}
\end{center}
\vspace{-0.7cm}
\caption{A slice of the nonlinear density field from the high-resolution 
    simulation with the RSD effect. The curvilinear grid shows the Eulerian 
    coordinate of each grid point of the potential isobaric coordinate.}
\label{fig:map_rsd}
\end{figure}

Figure \ref{fig:map} shows a slice of the nonlinear density through the 
high-resolution simulation, while the corresponding redshift space density
field is presented in Fig. \ref{fig:map_rsd}.
We also plot the Eulerian positions of a uniform grid of potential isobaric 
coordinates on the corresponding density field slice.
At small scales the random velocity dispersion elongates cosmic structures
along the line of sight, leading to the finger of God effect.
The finger of God effect smoothes the density fluctuations along the line of 
sight, blurring the clear small-scale structure in real space.
This causes the loss of cosmological information about the initial conditions.
Thus we expect that the performance in redshift space will be degraded compared 
to the result in real space.

The grid lines become closer in the higher density regions and sparser in the 
lower density regions. As a result, the mass enclosed in each curvilinear grid 
cell is approximately constant. Since the structure in redshift
space is blurred along the line of sight, the grid lines also move apart in the
dense regions compared to the corresponding real space grid line separation.
This suppresses the small-scale power of the reconstructed density field 
along the line of sight.
As the reconstructed density field is given by 
\bea
\delta_r^s(\bmxi)=-\nabla_{\bmxi}\cdot(\bms(\bmxi)-\bmxi)=3-\nabla_{\bmxi}\cdot\bms(\bmxi),
\eea
where $\nabla_{\bmxi}\cdot\bms(\bmxi)=\partial s_x/\partial\xi_x+\partial s_y/\partial\xi_y+\partial s_z/\partial\xi_z$, a larger separation $\Delta s_z$ 
between neighboring curvilinear grid lines implies a smaller density.
Although the squashing effect on large scales is not obvious in the density
field slice presented here, we still expect that the reconstructed density field
will have similar statistical properties as the redshift space nonlinear densiy
field but with much fewer nonlinearities.

\subsection{Correlation with the linear initial conditions}

To directly measure the linear initial conditions recovered with nonlinear 
reconstruction, we calculate the cross-correlation coefficient between the
density field and the linear initial conditions. 
The power spectra used for computing the cross correlation coefficients are 
averaged over ten simulations to reduce the cosmic variance.

\begin{figure}[tbp]
\begin{center}
\includegraphics[width=0.48\textwidth]{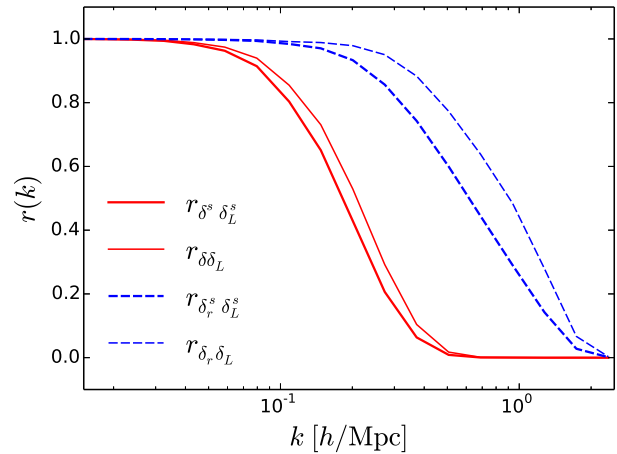}
\end{center}
\vspace{-0.7cm}
\caption{The cross-correlation coefficients with the initial conditions for the
    redshift space nonlinear (thick solid line) and reconstructed (thick dashed
    line) density fields and the nonlinear (thin sold line) and reconstructed 
    (thin dashed line) density fields without redshift space distortions.
    The correlation with the linear field for the redshift space field after 
    reconstruction is better than $50\%$ at $k\lesssim0.6\ h/\mr{Mpc}$ and 
    $80\%$ at $k\lesssim0.3\ h/\mr{Mpc}$.}
\label{fig:cc}
\end{figure}

We correlate the nonlinear density field $\delta(\bmk)$ and the reconstructed 
density field $\delta_r(\bmk)$ in real space with the linear density 
$\delta_L(\bmk)$.
For the redshift space density field $\delta^s(\bmk)$ and the reconstructed
density field $\delta_r^s(\bmk)$, we instead correlate them with the linear 
redshift space displacement divergence,
\bea
\delta^s_L(\bmk)=-i\bmk\cdot\bmp^s_L(\bmk)=(1+f\mu^2)\delta_L(\bmk),
\eea
where $\mu=k_z/k$.
To evaluate the overall performance, we first average the power spectrum over 
different directions and then compute the cross-correlation coefficient, i.e.,
the cross-correlation coefficient for the power spectrum monopole. 
Figure \ref{fig:cc} shows the correlation of the redshift space density field 
with the linear initial conditions before and after nonlinear reconstruction.
We also plot the cross-correlation coefficients for real space density fields
for comparison. 
The correlation with the linear field for the redshift space field after 
reconstruction is better than $50\%$ at $k\lesssim0.6\ h/\mr{Mpc}$ and $80\%$
at $k\lesssim0.3\ h/\mr{Mpc}$. For the redshift space nonlinear density field, 
the correlation is better than $50\%$ and $80\%$ only at $k\lesssim0.17\ h/\mr{Mpc}$ and $k\lesssim0.1\ h/\mr{Mpc}$.
Nonlinear reconstruction significantly reduces the shift nonlinearities and 
improves the number of linear modes by a factor $30$--$40$ depending on the 
scale used for comparison.

We note that RSD significantly degrade the performance of reconstruction.
However, the nonlinear density field is less affected by RSD; the correlation
with the linear density field is only slightly smaller than without RSD.
This is because RSD mainly scramble small-scale nonlinear structures and these 
small-scale structures have almost no correlation with the initial conditions.
However, we can recover the cosmological information which is still present in 
these structures. While RSD blur these visible small-scale structures, we can  
no longer recover the cosmological information that is present in the 
small-scale fluctuations of the real space density field.

\begin{figure}[tbp]
\begin{center}
\includegraphics[width=0.48\textwidth]{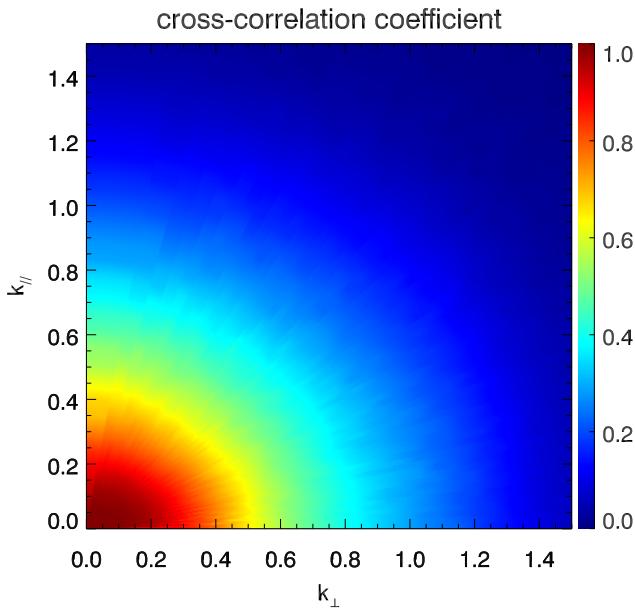}
\end{center}
\vspace{-0.7cm}
\caption{The two-dimensional cross-correlation coefficient of the reconstructed
density field $\delta_r^s(\bmk)$ with the redshift space linear field 
$\delta_L^s(\bmk)=(1+f\mu^2)\delta_L(\bmk)$.}
\label{fig:cc_Ls}
\end{figure}

As the solved displacement potential is anisotropic in the presence of RSD, we
expect that the performance of reconstruction will depend on the direction.
Figure \ref{fig:cc_Ls} shows the two-dimensional correlation coefficient $r_{\delta_r^s\delta_L^s}(k_\perp, k_\parallel)$ between the redshift space reconstructed density field and the linear density field. 
We also plot the correlation coefficients for three different $\mu$ bins,
$\mu=0,0.5,1$, in Fig. \ref{fig:cc_add}.
Note that the correlation with the linear field along the $k_\parallel$ axis
drops faster than along the $k_\perp$ axis.
One reason for this is that the Lagrangian velocity $\bmv(\bmq)$ is much more 
nonlinear than the Lagrangian displacement $\bmp(\bmq)$. The Lagrangian 
displacement is an integral of the Lagrangian velocity with respect to time,
\bea
\bmp(\bmq,t)=\int_0^t\bmv(\bmq,t')/a(t')dt'.
\eea
At higher redshifts the velocity field is more linear than the velocity field 
at later time. The Lagrangian displacement is a linear combination of velocities
at different redshifts with a different amount of nonlinearities, while the 
present-day Lagrangian velocity corresponds to the most nonlinear contribution
to the displacement. More details about this are presented in Appendixes
\ref{appendix:A} and \ref{appendix:B}.
Since the combination of $\nabla\cdot\bmp$ and $\nabla\cdot\bmv^s/aH$ is more
nonlinear than $\nabla\cdot\bmp$ and most power of $\nabla\cdot\bmv^s$ is 
concentrated in $\mu$ bins close to $\mu=1$, the correlation with the linear 
density field drops faster in these $\mu$ bins.

\subsection{Correlation with the nonlinear field}

\begin{figure}[tbp]
\begin{center}
\includegraphics[width=0.48\textwidth]{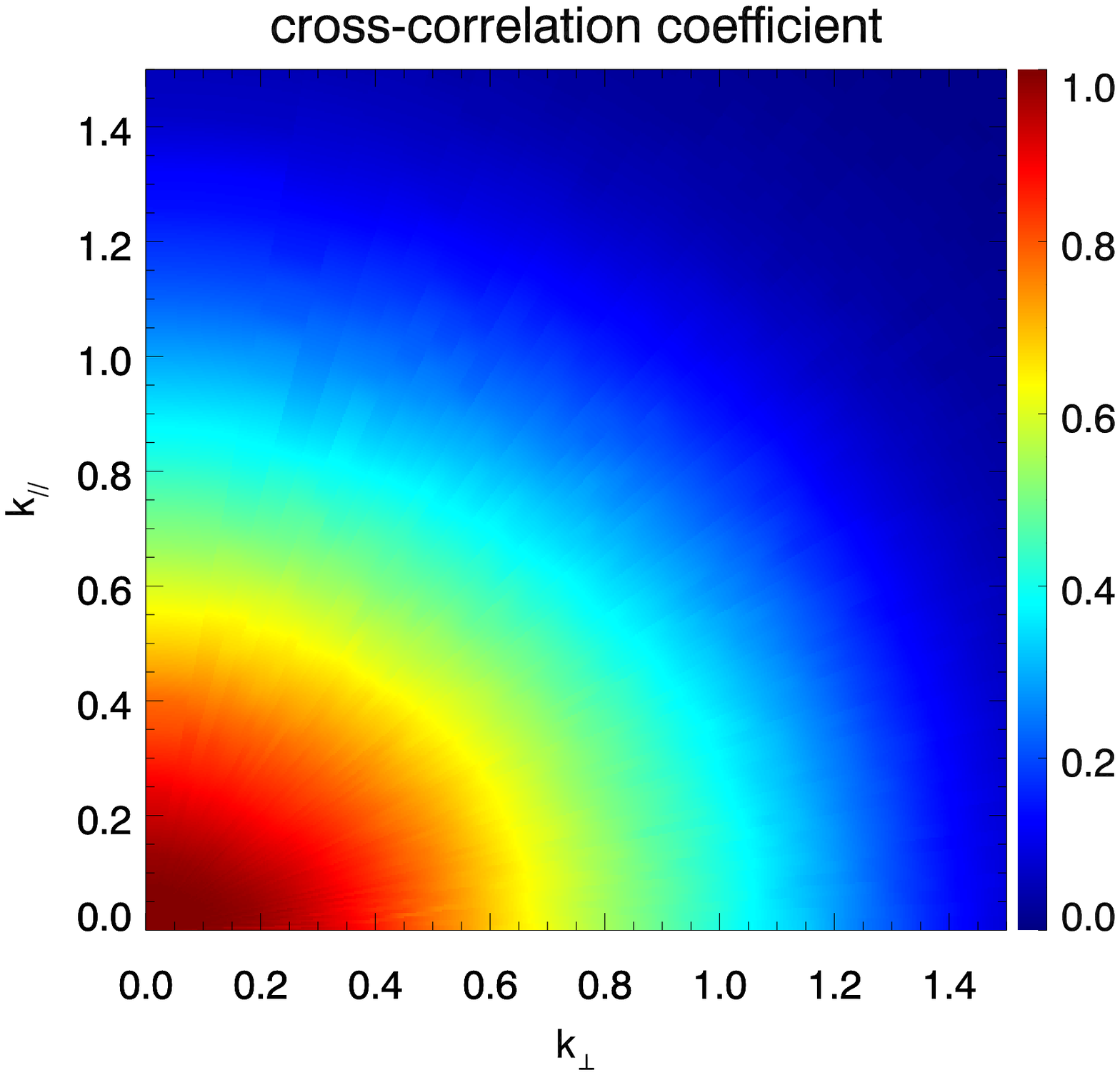}
\end{center}
\vspace{-0.7cm}
\caption{The two-dimensional cross-correlation coefficient of the reconstructed
density field $\delta_r^s(\bmk)$ with the redshift space nonlinear field 
$\delta_E^s(\bmk)=-i\bmk\cdot(\bmp+\bmv^s/aH)(\bmk)$.}
\label{fig:cc_Es}
\end{figure}

\begin{figure}[tbp]
\begin{center}
\includegraphics[width=0.48\textwidth]{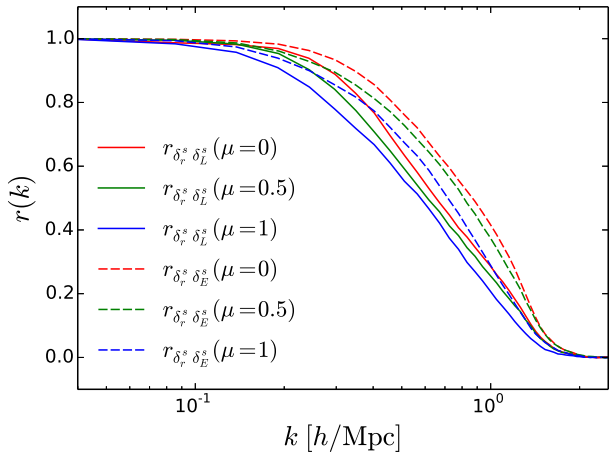}
\end{center}
\vspace{-0.7cm}
\caption{The correlation coefficients of the reconstructed density field 
    $\delta_r^s(\bmk)$ with the redshift space linear field $\delta_L^s(\bmk)$ 
    (solid lines) and the redshift space nonlinear field $\delta_E^s(\bmk)$ 
    (dashed lines) for three $\mu$ bins, $\mu=0,0.5,1$ (from right to left).}
\label{fig:cc_add}
\end{figure}

The correlation with the linear field quantifies the linear signal recovered
with nonlinear reconstruction, which is relevant for improving the measurements
of BAO scale in redshift surveys. 
To answer how well we reconstruct the nonlinear displacement, we need to
correlate the reconstructed density field with the redshift space nonlinear 
displacement divergence,
\bea
\delta_E^s(\bmk)=-i\bmk\cdot\bmp^s(\bmk)=-i\bmk\cdot(\bmp+\bmv^s/aH)(\bmk).
\eea
Figure \ref{fig:cc_Es} demonstrates the two-dimensional correlation coefficient 
$r_{\delta_r^s\delta_E^s}(k_\perp,k_\parallel)$ between the redshift space reconstructed density field and the redshift space nonlinear $E$-mode displacement.
We also plot the correlation coefficients for three different $\mu$ bins,
$\mu=0,0.5,1$, in Fig. \ref{fig:cc_add}.
The correlation with this nonlinear displacement field is much better than that 
with the linear displacement field as expected.
The redshift space nonlinear displacement divergence represents the best 
possible theoretical description of the reconstructed density field.

We find that the cross-correlation coefficient with the nonlinear field shows 
similar anisotropic feature as the cross-correlation coefficient with the 
linear field. This is because our reconstruction method cannot recover the 
$B$-mode displacement $\bmp_B+\bmv_B^s/aH$. The $B$-mode real space nonlinear 
displacement $\bmp_B$ is negligible on scales we care about. The $x$- and 
$y$-components of the $B$-mode shift velocity $\bmv^s_B$ are also negligible.
However, the $z$-component of the $B$-mode shift velocity $\bmv^s_B$ contributes
$8/35$ of the total power of $\bmv^s$ on large scales and even more on smaller
scales (see Appendix \ref{appendix:C}). 
Therefore, the correlation with the nonlinear field is slightly lower in the 
$\mu$ bins close to $\mu=1$ as shown in Fig. \ref{fig:cc_Es}.
The effect may also cause part of the degradation of the correlation coefficient
with the linear field along the line of sight.

\subsection{Reconstruction with transfer functions}

\begin{figure}[tbp]
\begin{center}
\includegraphics[width=0.48\textwidth]{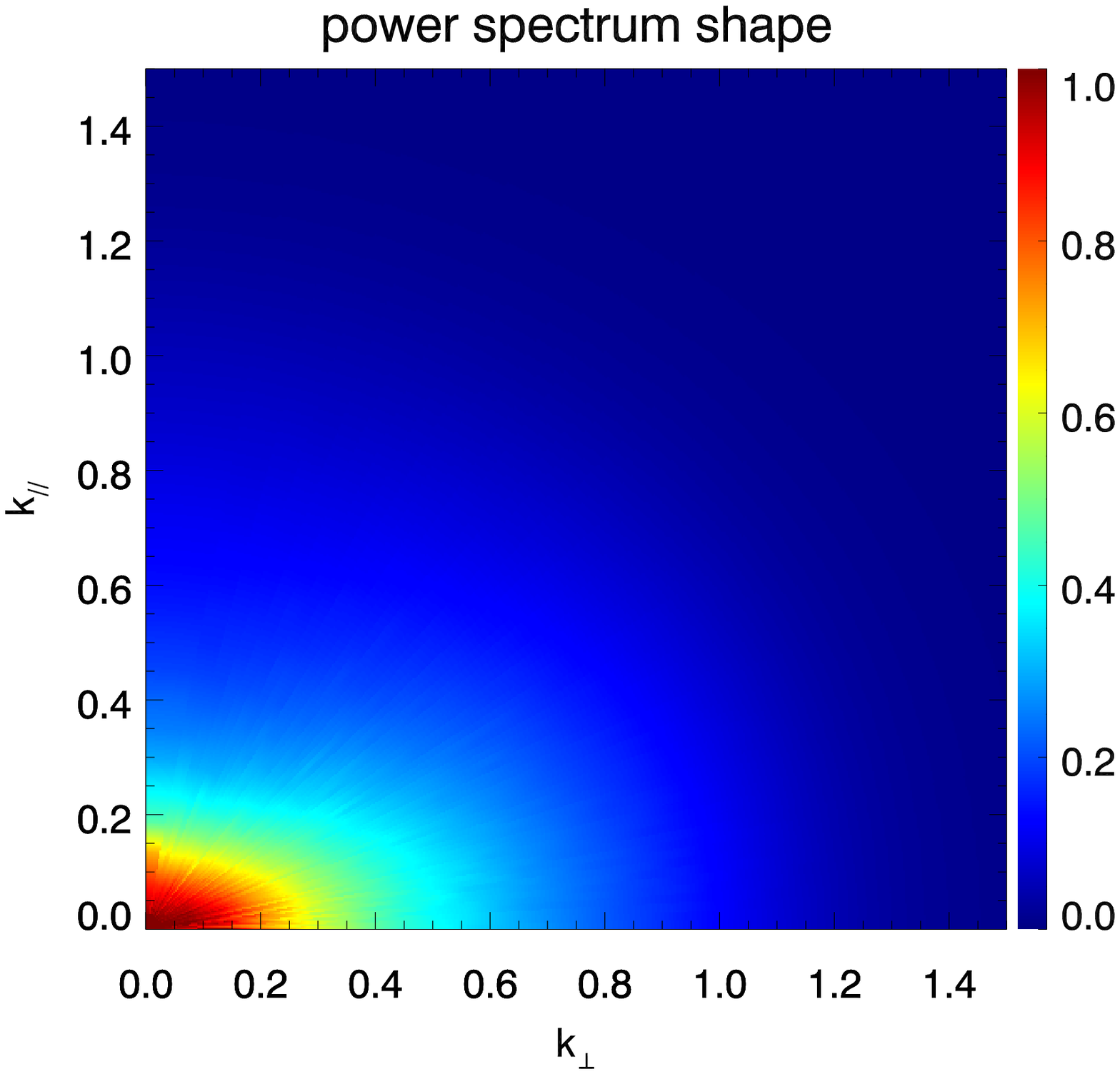}\\
\vspace{-0.3cm}
\includegraphics[width=0.48\textwidth]{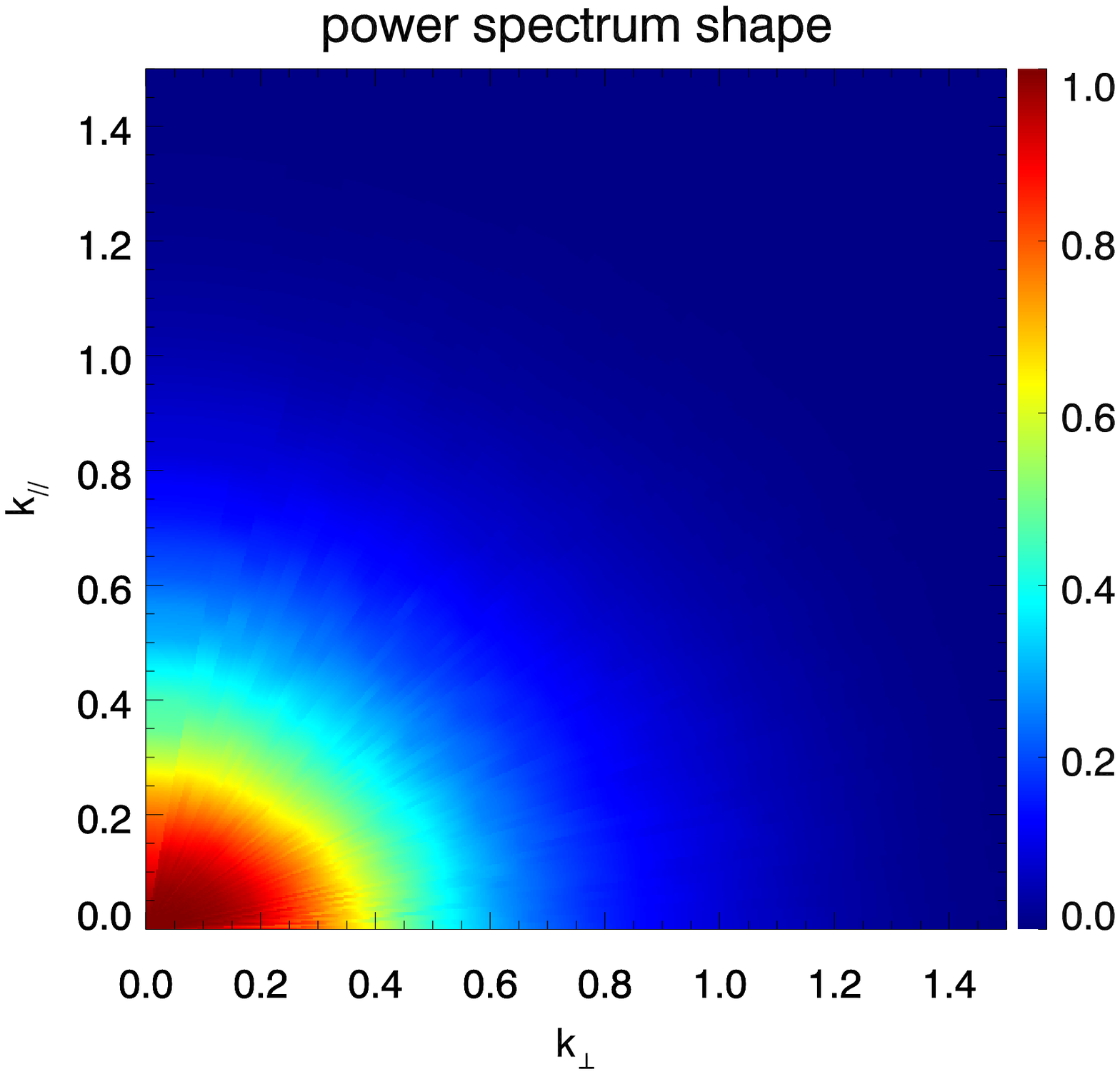}
\end{center}
\vspace{-0.7cm}
\caption{{\it Top:} The shape of the power spectrum for the reconstructed 
    density field normalized to the linear field, defined as 
    $P_{\delta_r^s}(\bmk)/P_{\delta_L^s}(\bmk)$.
    {\it Bottom:} The shape of the power spectrum for the filtered reconstructed density 
field normalized to the linear field, defined as 
$P_{\tilde{\delta}_r^s}(\bmk)/P_{\delta_L^s}(\bmk)$, where 
$\tilde{\delta}_r^s(\bmk)=t_L(\bmk)\delta_r^s(\bmk)$.}
\label{fig:shape_Ls}
\end{figure}

\begin{figure}[tbp]
\begin{center}
\includegraphics[width=0.48\textwidth]{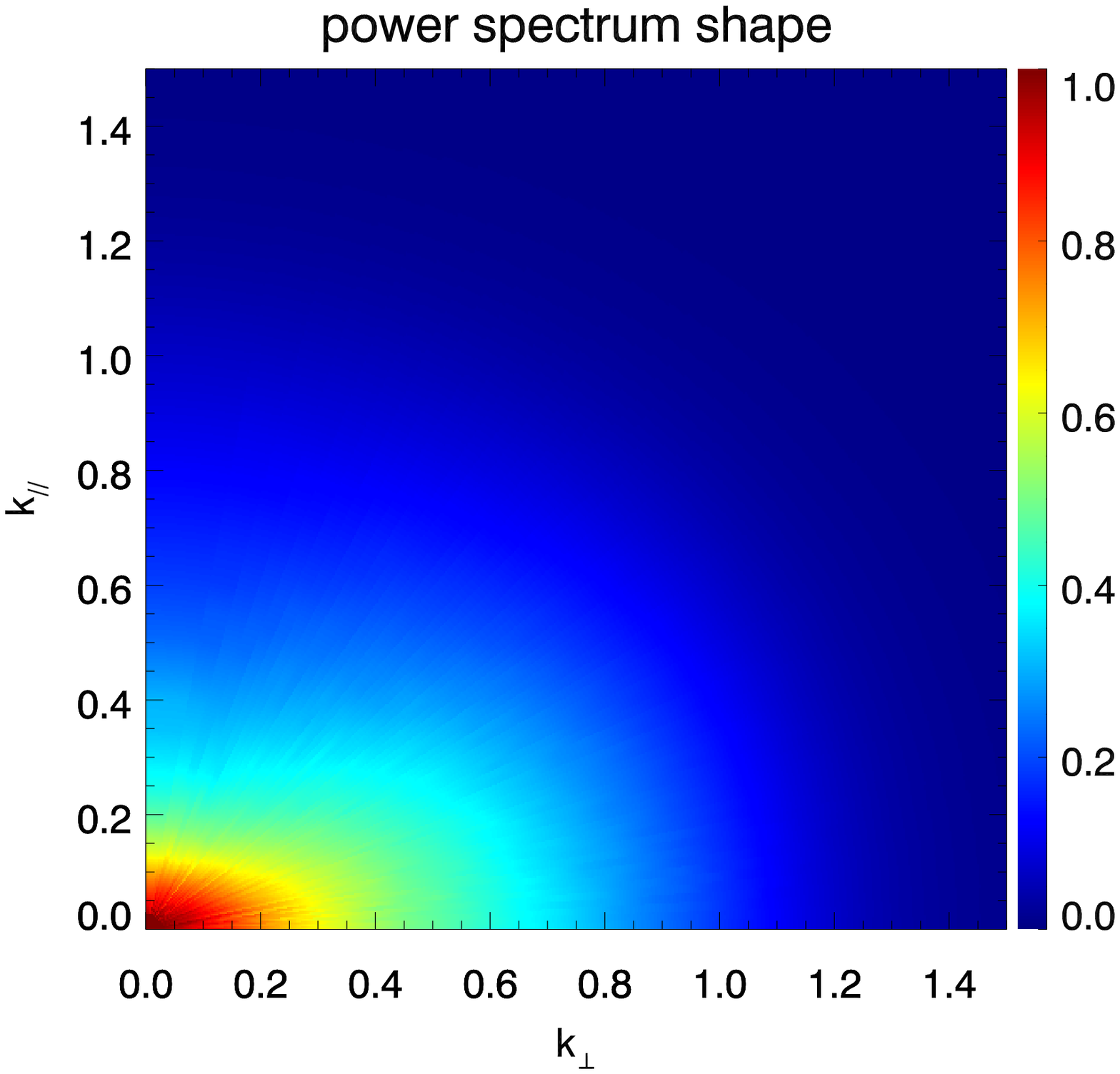}\\
\vspace{-0.3cm}
\includegraphics[width=0.48\textwidth]{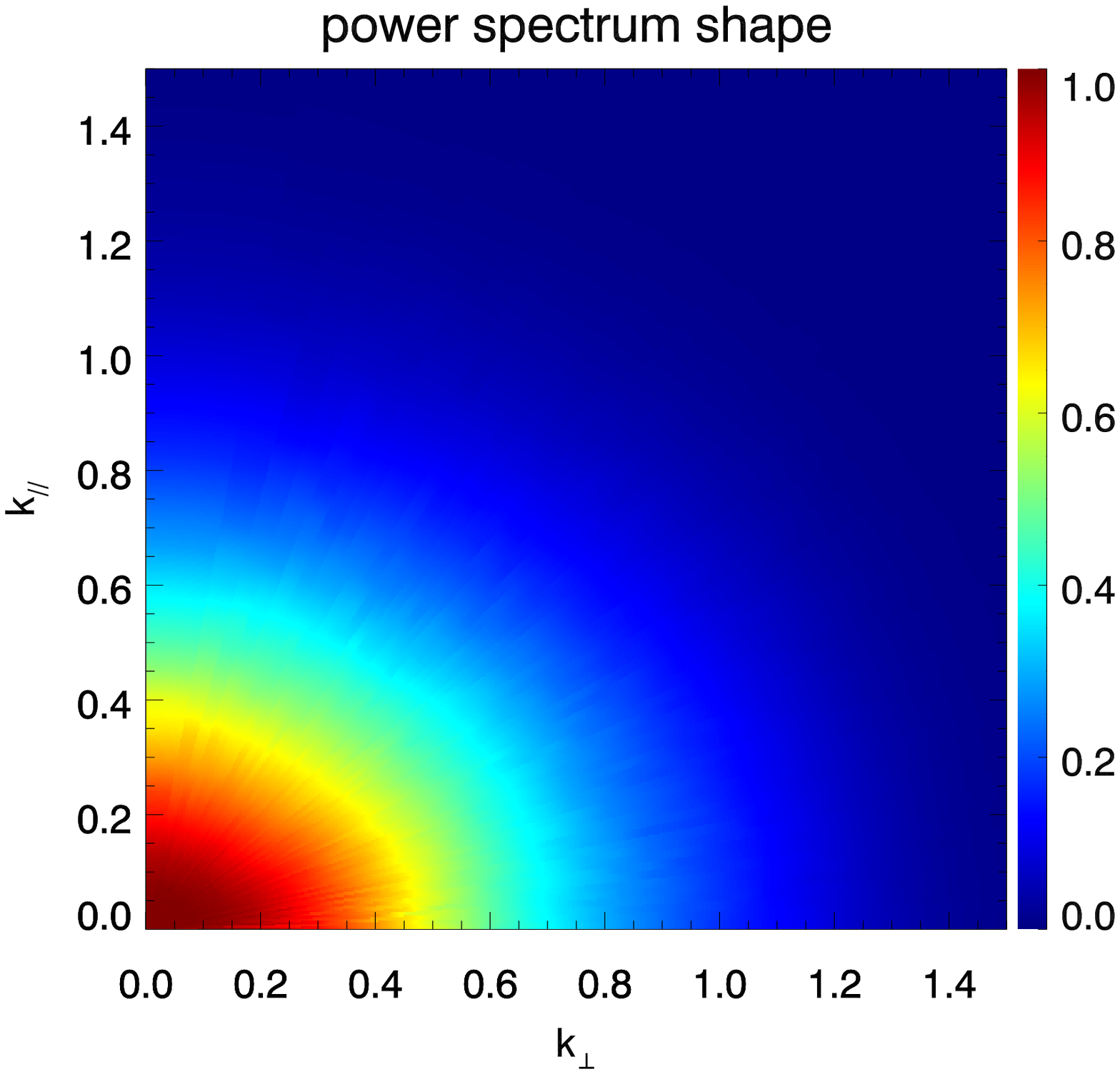}
\end{center}
\vspace{-0.7cm}
\caption{{\it Top:} The shape of the power spectrum for the reconstructed 
    density field normalized to the nonlinear field, defined as 
    $P_{\delta_r^s}(\bmk)/P_{\delta_E^s}(\bmk)$. 
    {\it Bottom:} The shape of the power spectrum for the filtered reconstructed density
field normalized to the nonlinear field, defined as
$P_{\tilde{\delta}_r^s}(\bmk)/P_{\delta_E^s}(\bmk)$, where 
$\tilde{\delta}_r^s(\bmk)=t_E(\bmk)\delta_r^s(\bmk)$.}
\label{fig:shape_Es}
\end{figure}

To recover the broadband power of the reconstructed density field, we can use 
the transfer functions \cite{2017Marcel}. We write the filtered reconstructed
field as
\bea
\tilde{\delta}_{r}^s(\bmk)=t(\bmk)\delta_r^s(\bmk),
\eea
where $t(\bmk)$ is the transfer function to be specified. 
Note that the transfer function is anisotropic in the presence of RSD. 
Here we consider the linear relation between different density fields and
neglect the higher order correlations.
We can choose the transfer function to minimize the difference between the 
reconstructed density field and the linear density field,
\bea
\la(t(\bmk){\delta}_r^s(\bmk)-\delta_L^s(\bmk))^2\ra,
\eea
and then the linear transfer function is given by
\bea
t_L(\bmk)=\frac{P_{\delta_r^s\delta_L^s}(\bmk)}{P_{\delta_r^s}(\bmk)},
\eea
where the subscript $L$ denoting the transfer function is to recover the broadband power of the reconstructed density field with respect to the linear field. 
We can also choose the transfer function to minimize the difference with respect
to the nonlinear field,
\bea
\la(t(\bmk)\delta_r^s(\bmk)-\delta_E^s(\bmk))^2\ra,
\eea
and then the nonlinear transfer function is given by
\bea
t_E(\bmk)=\frac{P_{\delta_r^s\delta_E^s}(\bmk)}{P_{\delta_r^s}(\bmk)},
\eea
where the subscript $E$ denoting the transfer function recovers the broadband
power spectrum with respect to the nonlinear field.

Figure \ref{fig:shape_Ls} shows the power spectra of the original and linear 
transfer function filtered reconstructed field, both normalized to the linear 
power spectrum.
The linear transfer function significantly improves the matching of the power 
spectrum of the reconstructed field to the linear power spectrum. 
For reconstruction with transfer function, the ratio between the two power spectra is larger than $0.5$ at $k\lesssim0.5\ h/\mr{Mpc}$ for $\mu$ bins around $\mu=0$ and $k\lesssim0.4\ h/\mr{Mpc}$ for $\mu$ bins around $\mu=1$. 
Again, we note a deficit in power along the line of
sight direction. This is because the shift velocity divergence 
$\nabla\cdot\bmv^s$ is more nonlinear than displacement and the reconstruction
misses the $B$-mode shift velocity $\bmv^s_B$ as we have discussed above.

Figure \ref{fig:shape_Es} shows the power spectra of the original and nonlinear
transfer function filtered reconstructed field, both normalized to the nonlinear
power spectrum.
The nonlinear transfer function significantly improves the matching of the power
spectrum of the reconstructed field to the nonlinear power spectrum. 
For reconstruction with transfer function, the ratio between the two power spectra is larger than $0.5$ at $k\lesssim0.6\ h/\mr{Mpc}$ for $\mu$ bins around $\mu\sim0$ and $k\lesssim0.5\ h/\mr{Mpc}$ for $\mu$ bins around $\mu\sim1$. 
Here the missed power along the line of sight is
solely due to the missed $B$-mode shift velocity $\bmv^s_B$.

\begin{figure}[tbp]
\begin{center}
\includegraphics[width=0.48\textwidth]{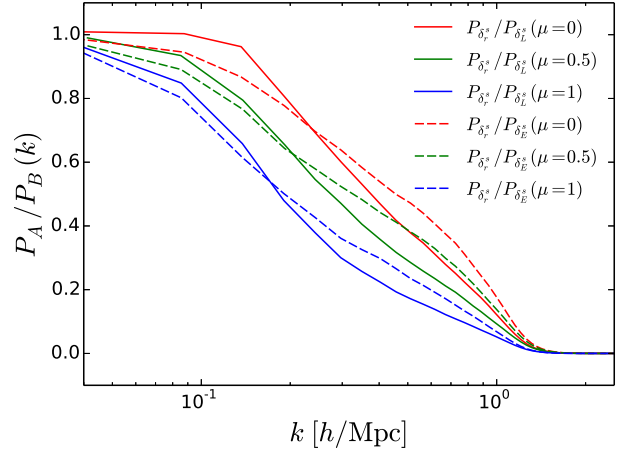}
\end{center}
\vspace{-0.7cm}
\caption{The reconstructed power spectrum divided by the linear (solid lines)
    and nonlinear (dashed) power spectra for three $\mu$ bins, $\mu=0,0.5,1$ 
    (from right to left).
}
\label{fig:shp_add}
\end{figure}

\begin{figure}[tbp]
\begin{center}
\includegraphics[width=0.48\textwidth]{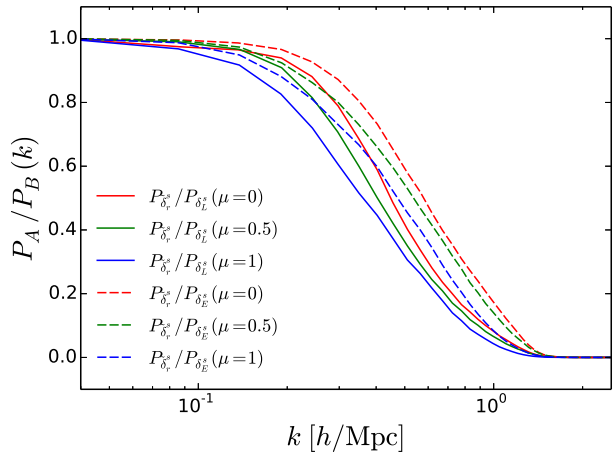}
\end{center}
\vspace{-0.7cm}
\caption{Same as Fig. \ref{fig:shp_add} but for the linear/nonlinear transfer
    function filtered reconstructed density fields. The solid lines show results
    for the reconstructed density field filtered with the linear transfer 
    function, i.e., $\tilde{\delta}_r^s(\bmk)=t_L(\bmk)\delta_r^s(\bmk)$, while
    the dashed lines show the result with the nonlinear transfer function, i.e.,
    $\tilde{\delta}_r^s(\bmk)=t_E(\bmk)\delta_r^s(\bmk)$.
}    
\label{fig:shptf_add}
\end{figure}

In Fig. \ref{fig:shp_add}, we plot the reconstructed power spectrum divided by 
the linear and nonlinear power spectra for three $\mu$ bins, $\mu=0,0.5,1$. 
We also present the corresponding results for the linear/nonlinear transfer 
function filtered fields in Fig. \ref{fig:shptf_add}. 
We notice that on relative large scales $k\lesssim0.1\ h/\mr{Mpc}$, the power 
spectrum amplitude ratio drops much faster than the correlation coefficient, which is also observed in other reconstruction methods \cite{2017Marcel,2017US}.
Recovering the cross correlation is easier than the full broadband power spectrum.
This problem can be solved by using the transfer function as we indeed observe
in Fig. \ref{fig:shptf_add}.

\begin{figure}[tbp]
\begin{center}
\includegraphics[width=0.48\textwidth]{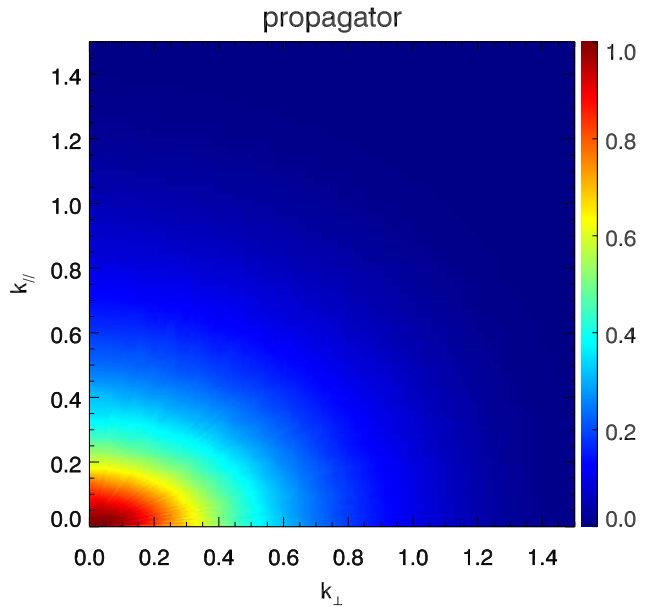}
\end{center}
\vspace{-0.7cm}
\caption{The propagator $C_L(\bmk)$ for the linear field. The additional damping
along the line of sight is caused by the nonlinearity of the shift velocity and the finger of God effect.}
\label{fig:tf_Ls}
\end{figure}

\begin{figure}[tbp]
\begin{center}
\includegraphics[width=0.48\textwidth]{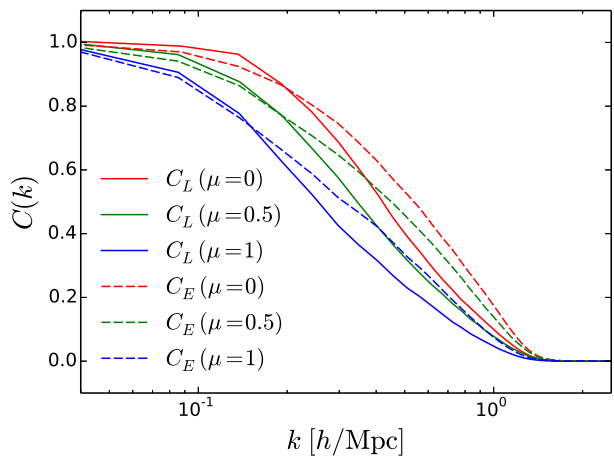}
\end{center}
\vspace{-0.7cm}
\caption{The linear propagator $C_L(\bmk)$ (solid lines) and the nonlinear 
    propagator $C_E(\bmk)$ (dashed lines) for three $\mu$ bins, $\mu=0,0.5,1$ 
    (from right to left).
}
\label{fig:prop_add}
\end{figure}

The transfer function introduced above is similar to the Wiener filter. 
The reconstructed density field can be written as 
\bea
\delta_r^s(\bmk)=C_L(\bmk)\delta_L^s(\bmk)+N_L(\bmk),
\eea
where $C_L(\bmk)=P_{\delta_r^s\delta_L^s}(\bmk)/P_{\delta_L^s}(\bmk)$ is the 
propagator and $N_L(\bmk)$ is the noise with respect to the linear field. 
The Wiener filter for the linear signal in reconstructed density field is 
\bea
W_L(\bmk)=\frac{C^2_L(\bmk)P_{\delta_L^s}(\bmk)}{C_L^2(\bmk)P_{\delta^s_L}(\bmk)+P_{N_L}(\bmk)},
\eea
which differs from the linear transfer function $t_L(\bmk)$ by a propagator 
$C_L(\bmk)$,
\bea
t_L(\bmk)=\frac{W_L(\bmk)}{C_L(\bmk)}.
\eea
The transfer function also deconvolves the propagator from the density field 
in addition to applying the Wiener filter to recover the power spectrum shape.
Figure \ref{fig:tf_Ls} shows the two-dimensional propagator with respect to the
linear field. We also plot the propagator with respect to the linear field for 
three $\mu$ bins, $\mu=0,0.5,1$, in Fig. \ref{fig:prop_add}.
The propagator shows additional suppression along the $z$-direction. 
This is caused by the nonlinearity of the shift velocity $\bmv^s$ and the finger of God effect. Since the combination of displacement divergence $\nabla\cdot\bmp$ and velocity divergence $\nabla\cdot\bmv^s$ is more nonlinear than $\nabla\cdot\bmp$ itself and most power of the shift velocity divergence $\nabla\cdot\bmv^s$ is concentrated around $\mu\sim1$, the $\mu$ bins close to $\mu=1$ contain
less information about the linear initial conditions.
The small-scale random velocity dispersion suppresses the density fluctuations
along the line of sight, leading to larger separations along the $z$-axis 
between neighboring grid lines. 
The nonlinear effects due to random motion on small scales not only cause the 
suppression of power along the line of sight, but also scramble the information
about linear initial conditions along this direction. 

\begin{figure}[tbp]
\begin{center}
\includegraphics[width=0.48\textwidth]{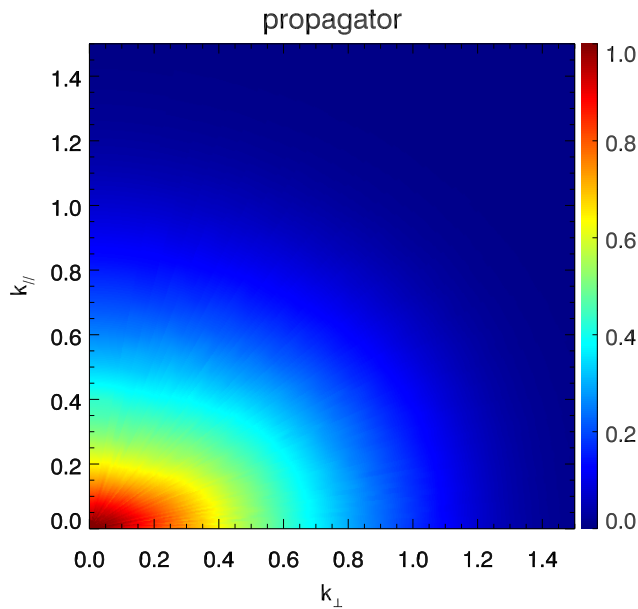}
\end{center}
\vspace{-0.7cm}
\caption{The propagator $C_E(\bmk)$ for the nonlinear field. The additional 
damping along the line of sight is caused by the finger of God effect only as 
we are correlating with the nonlinear field.}
\label{fig:tf_Es}
\end{figure}

We can also write the reconstructed density field as 
\bea
\delta_r^s(\bmk)=C_E(\bmk)\delta_E^s(\bmk)+N_E(\bmk),
\eea
where $C_E(\bmk)=P_{\delta_r^s\delta_E^s(\bmk)}/P_{\delta_E^s}(\bmk)$ and 
$N_E(\bmk)$ are the propagator and noise with respect to the nonlinear field.
The similar relation $t_E(\bmk)={W_E(\bmk)}/{C_E(\bmk)}$ also holds for the 
Wiener filter for the nonlinear signal 
\bea
W_E(\bmk)=\frac{C^2_E(\bmk)P_{\delta_E^s}(\bmk)}{C_E^2(\bmk)P_{\delta^s_E}(\bmk)+P_{N_E}(\bmk)}.
\eea
Figure \ref{fig:tf_Es} shows the two-dimensional propagator with respect to the
nonlinear field. We also plot the propagator with respect to the nonlinear field
for three $\mu$ bins, $\mu=0,0.5,1$ in Fig. \ref{fig:prop_add}.
Here the additional damping along the line of sight is only due to the Finger 
of God effect as we are correlating with the nonlinear field.

To extract distinct features like BAO wiggles from the power spectrum of the
reconstructed density field, we can directly apply the transfer function 
calibrated from simulations or mocks. 
In the BAO-only analysis, the transfer function is only used to match the power 
spectrum amplitude with the linear density field and itself does not contain any
information about the BAO wiggles if calibrated to simulations with 
no-wiggle initial linear power spectra \cite{2017Marcel}.
However, if we want to extract additional cosmological information about RSD 
from the density field, we need to model the broadband power spectrum of the
reconstructed density field. 
The linear power spectrum $P_{\delta_L^s}(\bmk)$ or the nonlinear power $P_{\delta_E^s}(\bmk)$ does not suffice to model the reconstructed density field, as 
they only describe the coherent part of the reconstructed density field. 
We need to consider the small-scale random velocity dispersion and the errors of
reconstruction, i.e., the propagator $C(\bmk)$ and the noise term $P_N(\bmk)$.
For the linear field, we also need to consider nonlinearities of the coherent 
part of the reconstructed field for modeling the mode-coupling term $P_N(\bmk)$.
Probably these can be studied using higher order Lagrangian perturbation theories or the effective field theories of large-scale structure \cite{2016BSZ}.
Similar works for the standard reconstruction method have been conducted \cite{2009PWC,2009NWP,2015marcel,2015Martin,2016Seo,2017CKA}.
We can directly calibrate them against simulations or mock catalogs but the
computation cost would be expensive. We leave this for future work.

\subsection{Correlation with displacements and velocities}

\begin{figure}[tbp]
\begin{center}
\includegraphics[width=0.48\textwidth]{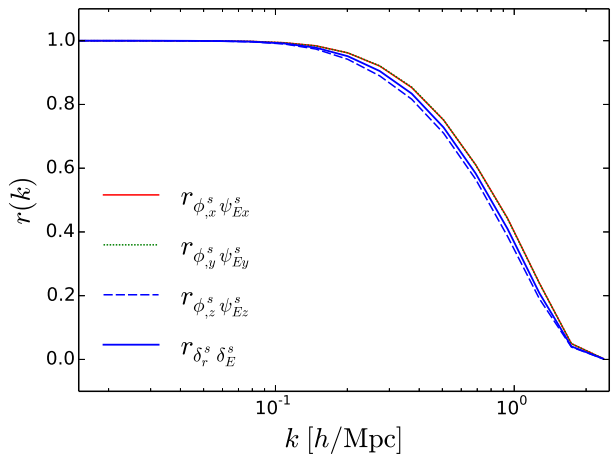}
\end{center}
\vspace{-0.7cm}
\caption{The cross-correlation coefficients of the reconstructed displacement
    components with the nonlinear displacement components (solid, dotted, dashed
    lines are for $x,y,z$-components respectively) and the reconstructed
density field with the nonlinear displacement divergence.}
\label{fig:cc_disp}
\end{figure}

As we have reversed the nonlinear mapping from the initial Lagrangian position
to the final redshift space position up to shell crossing during nonlinear 
reconstruction, the gradient of the solved displacement potential 
\bea
\nabla\phi^s=(\phi_{,x}^s,\phi_{,y}^s,\phi_{,z}^s)
\eea
gives an estimate of the nonlinear displacement
\bea
\bmp_E^s=(\psi_{Ex}+\frac{v_{Ex}^s}{aH},\psi_{Ey}+\frac{v_{Ey}^s}{aH},\psi_{Ez}+\frac{v_{Ez}^s}{aH}).
\eea
To examine the performance of reconstruction along different directions, we 
compute the correlation coefficient for each component separately. 
Figure \ref{fig:cc_disp} shows the cross-correlation coefficients. 
For comparison we also plot the cross-correlation coefficient between the 
reconstructed density field and the nonlinear displacement divergence.
Here we first average the power spectrum over different $k$-space directions
and then compute the cross-correlation coefficient. 
The cross correlation for the $x$- and $y$-component of the displacement is
better than the $z$-component as expected, since the missed $z$-component 
$v_{Bz}^s$ of the $B$-mode shift velocity is much larger than $v_{Bx}^s$ and 
$v_{By}^s$. Though we cannot reconstruct these $B$-mode shift velocities, they
still change the observed mass distribution used for nonlinear reconstruction; 
as a result, they induce noises for the reconstructed displacement 
respectively, reducing the correlations with the real $E$-mode displacements. 

\begin{figure}[tbp]
\begin{center}
\includegraphics[width=0.48\textwidth]{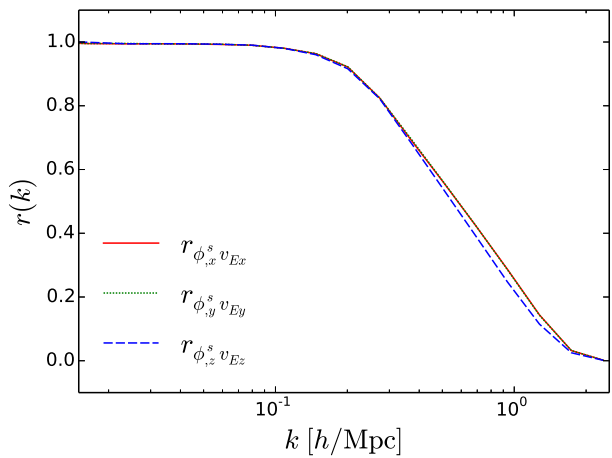}
\end{center}
\vspace{-0.7cm}
\caption{The cross-correlation coefficient of the reconstructed displacement
with the Lagrangian velocity. The correlation for each component is larger
than $0.5$ at $k\lesssim0.43\ h/\mr{Mpc}$, much smaller than the scale where 
the nonlinear density field correlates with the linear density field.}
\label{fig:cc_vel}
\end{figure}

In linear theory, the Lagrangian displacement only differs from the Lagrangian 
velocity by a factor $afH$. Their relation becomes complicated in higher order
Lagrangian perturbation theory. 
Nevertheless, it is still interesting to study the cross correlation of the 
reconstructed displacement with the real Lagrangian velocity.
Figure \ref{fig:cc_vel} shows the cross-correlation coefficients computed using
the angular-averaged power spectra.
The cross-correlation coefficient for each component is larger than $0.5$ at
$k\lesssim0.43\ h/\mr{Mpc}$, though the behavior of the $z$-direction is a 
little different from the other two directions.
This scale is much smaller than the scale where the nonlinear density field 
correlates with the linear density field ($k\lesssim0.17\ h/\mr{Mpc}$).
Therefore, we expect that the new velocity reconstruction scheme based on the 
relation between the reconstructed displacement and the Lagrangian velocity
would be better than the method based on the linearized continuity equation.

\section{Applications}
\label{sec:app}

The prime application of nonlinear reconstruction is to reverse the large-scale
bulk flows and improve the precision measurement of the BAO scale. 
The density field reconstructed with the dark matter density in real space has
a comparable fidelity as the linear density field for measuring the BAO scale,
since the BAO wiggles are washed out at $k>0.6\ h/\mr{Mpc}$ in both fields.
The new reconstruction methods can restore the BAO signal of the linear density
perfectly in the idealized situation \cite{2017Marcel,2017Wang}.
The shot noise limits the performance of reconstruction when we apply 
nonlinear reconstruction to real space halo fields \cite{2017Yu}.
The new reconstruction method requires a higher number density
$\bar{n}\geq10^{-3}\ (h/\mr{Mpc})^3$ to have a better performance than the 
standard reconstruction method \cite{2017Yu}.
In this paper we show that RSD further degrade the performance due to the 
nonlinearity of the shift velocity and the small-scale random velocity 
dispersion.
Nevertheless, we still expect the new method to substantially improve the 
precision of the measured BAO scale of dense low redshift large-scale 
structure surveys, including SDSS main sample, DESI BGS, and 21 cm intensity
mapping, while the improvement over standard reconstruction for
less dense high redshift samples like BOSS LOWZ and CMASS will be moderate.

We have applied reconstruction to halo density fields in redshift space. 
The two-dimensional correlation coefficients of the reconstructed density field
with the linear initial conditions are presented in Appendix \ref{appendix:D}.
To apply nonlinear reconstruction to observations, we need to consider several 
complicated effects. A practical issue related to this is the boundary of a 
survey region, which is not the periodic simulation box. 
This problem can be solved by embedding the observed galaxy density field 
into a uniform background. We have demonstrated a simple test about this 
solution in Appendix \ref{appendix:E}. 
The practical application of nonlinear reconstruction to SDSS galaxy samples 
will be studied in a future paper.

The most innovative result of this paper is that the nonlinear reconstruction
method can improve not only the BAO measurement but also the RSD measurement.
The nonlinear mapping solved from the anisotropic redshift space density field
includes the $E$-mode shift velocity $\bm{v}^s_E$, which contains cosmological
information about the growth of structure. 
The reconstructed density field shares many similar features as the redshift 
space density field like the large-scale squashing effect and small-scale 
finger of God effect, but much more linear than the original redshift space 
density field. A large amount of shift nonlinearities has been removed in the
nonlinear reconstruction. The number of linear modes that can be used to measure the growth rate is increased by about $30$--$40$ times after reconstruction. 
Therefore, we expect that this will significantly reduce uncertainties of the 
measured structure growth rate since the constraining power of galaxy surveys
scales steeply with the number of modes included in the RSD analysis.
In contrast to the BAO reconstruction, where nonlinear reconstruction still 
needs a higher number density to gain better performance than standard 
reconstruction, the RSD nonlinear reconstruction can always obtain better 
performance than without reconstruction, no matter the number density of galaxy 
samples. This has significant implications for constraining modified gravity
theories using the measured linear structure growth rate.
However, we still expect substantial improvements for the dense low-redshift 
surveys as discussed above while the performance for less dense samples
will be limited by the shot noise. 

The current BAO reconstruction involves displacing particles according to the 
large-scale linear displacement field computed from the observed galaxy density
field with some certain model assumptions like the smoothing scale, galaxy bias,
growth rate etc \cite{2007bao}. The results depend on the assumed fiducial
model and have to be tested against different parameter choices, which are 
computationally expensive \cite{2012nikhil}.
The new reconstruction method directly solves the displacement potential from 
the observed density field, which is a purely mathematical problem without any
cosmological dynamics involved.
The parameters like the halo bias and structure growth rate do affect the 
reconstructed density field, but they are to be jointly fitted when we model 
the power spectrum after reconstruction, while there are no cosmological 
parameters involved in the reconstruction process itself.
This saves a large amount of computational costs when applying reconstruction
to galaxy surveys to improve BAO and RSD measurements.

The fact that the reconstructed displacements correlate with the Lagrangian 
velocities to nonlinear scales in Lagrangian space implies that we can construct
new peculiar velocity reconstruction schemes based on this relation.
The current peculiar velocity reconstruction methods are often based on the 
linearized continuity equation like the one used to measure the kinematic 
Sunyaev-Zeldovich effect through cross correlation \cite{2016PVR}. 
The linearized continuity equation is 
\bea
\dot{\delta}+\nabla\cdot\bmv=0.
\eea
Assuming the velocity is curl free, the velocity can be inferred from the 
density through
\bea
\bmv(\bmk)=afH\frac{i\bmk}{k^2}\delta(\bmk)W_R(k),
\eea
where $W_R(k)$ is a smoothing function with the characteristic scale $R$.
The smoothing scale should be chosen close to the scale where linear theory 
breaks down \cite{2015marcel,2016Seo}. 
To make the above linear approximation valid, the nonlinear density field 
usually needs to be smoothed on scales $\sim10\ \mr{Mpc}/h$ to remove small-scale nonlinearities. This is because all quantities defined in Eulerian space
like $\delta(\bmx)$ and $\bmv(\bmx)$ involve very strong shift nonlinearities,
which have to be reduced in order to apply the linear relation. 
However, a lot of cosmological information about small-scale velocities is lost 
in this process. The quantities defined in Lagrangian space like Lagrangian 
displacement $\bmp(\bmq)$ and Lagrangian velocity $\bmv(\bmq)$ are free of any
nonlinear shift terms \cite{2016BSZ}.
This motivates us to develop new velocity reconstruction methods using the 
relation between Lagrangian quantities instead of Eulerian quantities.

In the Lagrangian perturbation theory, the Lagrangian displacement is solved as
\bea
\bmp(\bmq,t)=\bmp^{(1)}(\bmq,t)+\bmp^{(2)}(\bmq,t)+\bmp^{(3)}(\bmq,t)+\cdots,
\eea
where $\bmp^{(n)}\propto (\bmp^{(1)})^n$ is the $n$th-order solution. 
The Lagrangian velocity of mass element with initial coordinate $\bmq$ is the 
time derivative of the Lagrangian displacement 
\bea
\bmv(\bmq,t)=a\dot{\bmx}(\bmq,t)=a\dot\bmp(\bmq,t).
\eea
The time dependence of the $n$th-order solution $\bmp^{(n)}$ can be well
approximated as $\bmp^{(n)}\propto D^n$. Therefore, the Lagrangian velocity
can be expressed as
\bea
\bmv(\bmq,t)=afH\bmp^{(1)}(\bmq,t)+2afH\bmp^{(2)}(\bmq,t)+\cdots.
\eea
The most direct and simple idea is to filter the reconstructed displacement
field with the linear transfer function to minimize the difference with the
linear displacement field. Then compute the Lagrangian velocity using the linear
relation $\bmv^{(1)}(\bmq)=afH\bmp^{(1)}(\bmq)$ and next generate the Eulerian space velocity field using a velocity assignment method \cite{2015NP1,2015NP2}.
This $\mathcal{O}(1)$ velocity reconstruction is the Lagrangian space version 
of the current velocity reconstruction in Eulerian space. However, since much
fewer shift nonlinearities are involved in Lagrangian space and the displacement
and velocity correlate to a much smaller scale in Lagrangian space than the
density and velocity in Eulerian space, we expect that the new method will have
a better performance than the previous methods operated in Eulerian space.
We can also compute the second-order solution $2afH\bmp^{(2)}$ using the 
linear transfer function filtered displacement since the displacement is much
more linear than the Lagrangian velocity. Then we can use the second-order 
transfer functions to optimally combine $afH\bmp^{(1)}$ and $2afH\bmp^{(2)}$
to obtain the reconstructed Lagrangian velocity \cite{2017Marcel}.
The $\mathcal{O}(2)$ velocity reconstruction scheme is expected to better 
recover small-scale velocities. 
One more possibility is to directly correlate the reconstructed displacement 
with the nonlinear Lagrangian velocity and solve the optimal transfer function.
The detailed tests of the new velocity reconstruction schemes and comparisons 
against the current velocity reconstruction method will be investigated in a 
future paper.

\section{Conclusions}
\label{sec:cls}

In this paper we apply nonlinear reconstruction to redshift space nonlinear
dark matter density field and reconstruct the displacement potential that 
describes the mapping from initial Lagrangian coordinates to final redshift
space coordinates. We undo the shifts from Lagrangian positions to 
redshift space positions by mapping the nonlinear density field into a uniform
field and reduce the associated nonlinearities significantly.
The reconstructed density field is defined as the negative divergence of the reconstructed displacement or the negative Laplacian of the displacement potential.
The cross-correlation coefficient between the reconstructed density field and 
the linear initial conditions is larger than $0.5$ at $k\lesssim0.6 h/\mr{Mpc}$
and $0.8$ at $k\lesssim0.3\ h/\mr{Mpc}$. We note that the reconstructed density
field shares similar properties with the redshift space nonlinear density field,
including the nonlinear finger of God effect and linear Kaiser effect. 
However, the reconstructed density field involves much fewer nonlinearities 
related with the shifts from Lagrangian positions to redshift space positions.
Therefore we expect this nonlinear reconstruction approach to substantially
improve the cosmological information content of BAO and RSD for dense 
large-scale structure surveys.

Reconstruction of the nonlinear displacement field is not perfect and limited 
by several factors, including shell crossing, vorticity and grid limiters.
The reconstruction is not sensitive to the parameters used for grid limiters 
like the maximal compression factor and expansion volume limit (see tests presented in Ref. \cite{2016HMZ}). The performance is mostly limited by shell crossing
and vorticity. Note that the $B$-mode displacement will also change the mass 
distribution in the Universe in addition to the $E$-mode displacement. 
However, due to the limited degree of freedom, we can only reconstruct a scalar
field (displacement potential) from the observed density field which is caused 
by both $E$- and $B$-mode displacements; as a result, this induces errors 
in the reconstructed displacement field. The mapping from Lagrangian coordinates
to redshift space coordinates is not unique once shell crossing occurs. 
We can still define a coordinate system which we refer to as potential isobaric
gauge/coordinates where the mass per volume element is approximately constant.
However, the reconstructed displacement is not accurate in the regime where 
shell crossing occurs. The shell crossing induces a similar effect like the 
vorticity; they both cause changes in the mass distribution in the Universe,
but we treat the observed density field as a field not affected by them.
To disentangle these two effects, we need to consider reconstruction in the 
one-dimensional case, where the displacement field has no $B$-mode component.
By performing reconstruction in the one-dimensional Universe, we can estimate 
the effect solely induced by shell crossing \cite{2016HMZ1}. 

In addition to the effects discussed above, the correlation of the reconstructed
density with the linear initial conditions is also affected by the inherent
nonlinearities of the Lagrangian displacement and velocity. 
Although Lagrangian quantities like the displacement and velocity are free of 
shift terms, they are still affected by structure formation, which induces
inherent nonlinearities \cite{2012TZ}. 
These inherent nonlinearities cannot be removed by nonlinear reconstruction. 
In summary, the performance of initial condition reconstruction is limited by
residual shift nonlinearities due to imperfect reconstruction of the nonlinear
displacement field and nonlinearities of the displacement and velocity fields.

The nonlinear reconstruction approach provides a new route to measure the RSD
effect. Instead of modeling the complicated mapping from real to redshift space,
we can reconstruct the Lagrangian displacement and velocity directly and use
the reconstructed density field to measure the anisotropic effects due to the
peculiar velocity. When trying to model RSD, we are creating nonlinear effects 
to make the theoretical power spectrum look like the nonlinear power spectrum.
This still does not improve the information in the power spectrum of the 
observed density field from surveys. 
However, reconstruction can reduce nonlinearities and reverse the mapping from 
Lagrangian coordinates to redshift space coordinates.
The density field after reconstruction can be described using linear theory on 
scales where the redshift space density field already needs to be modeled using
higher order perturbation theory.

The correlation of the reconstructed displacement with the Lagrangian velocity
is much better than the correlation between density and velocity in Eulerian 
space. This implies that the new peculiar velocity reconstruction scheme can be
constructed based on the relation between Lagrangian displacement and velocity.
Since much fewer nonlinearities are involved in the new reconstruction schemes,
we expect that the performance will be better than the Eulerian velocity
reconstruction method. We propose several velocity reconstruction methods here 
and detailed tests will be presented in future.

In this paper we show that the nonlinear reconstruction not only works in real 
space but also can be implemented in redshift space. 
In addition to improving the linear BAO signals, we find that this nonlinear
approach can also improve the RSD measurements since the reconstructed density
field shares similar features as the redshift space nonlinear density field and 
much more linear modes are recovered after reconstruction. 
This could substantially recover cosmological information about BAO and RSD for
dense low-redshift large structure surveys such as SDSS MGS, DESI BGS and 21 cm
intensity mapping surveys.


\section*{acknowledgements}
We thank Pengjie Zhang and Gong-Bo Zhao for valuable discussions.
We acknowledge the support of the Chinese Ministry of Science and Technology 
under Grant No. 2016YFE0100300,
the National Natural Science Foundation of China under Grants No. 11633004, 
No. 11373030, No. 11773048 and No. 11403071, CAS Grant No. QYZDJ-SSW-SLH017,
and Natural Sciences and Engineering Research Council of Canada.
The simulations are performed on the BGQ supercomputer at the SciNet HPC
Consortium. SciNet is funded by the following: the Canada Foundation for 
Innovation under the auspices of Compute Canada, the Government of Ontario, 
Ontario Research Fund---Research Excellence, and the University of Toronto.
The Dunlap Institute is funded through an endowment established by the David 
Dunlap family and the University of Toronto.
Research at the Perimeter Institute is supported by the Government of Canada
through Industry Canada and by the Province of Ontario through the Ministry of
Research and Innovation.

\appendix

\section{Lagrangian displacement field}
\label{appendix:A}

\begin{figure}[tbp]
\begin{center}
\includegraphics[width=0.48\textwidth]{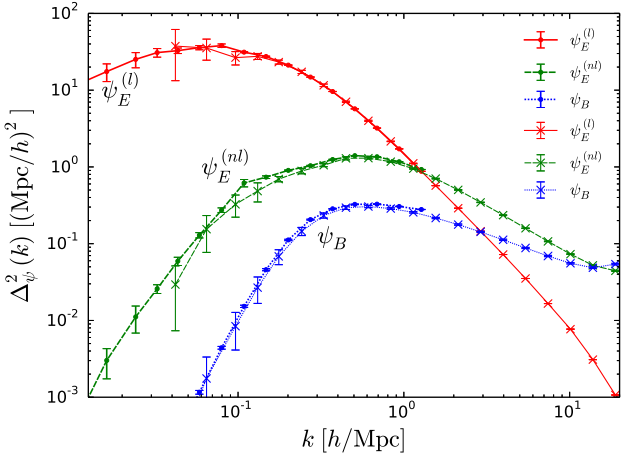}
\end{center}
\vspace{-0.7cm}
\caption{The power spectra for the decomposed Lagrangian displacements.
The large-scale nonlinear displacement field is well described by the Zeldovich
approximation for $k\lesssim1\ h/\mr{Mpc}$. The nonlinear $E$-mode displacement
becomes important at $k\gtrsim1\ h/\mr{Mpc}$. The $B$-mode displacement only 
matters on very small scales for $k\gtrsim10\ h/\mr{Mpc}$. The power spectra 
are computed using two sets of simulations with different resolutions.}
\label{fig:pk_disp}
\end{figure}

Any vector field can be separated into a gradient part and a curl part. 
Hence we can decompose the real space nonlinear displacement as
\bea
\bmp(\bmq)=\bmp_E(\bmq)+\bmp_B(\bmq),
\eea
where $\bmp_E=(\bmk\cdot\bmp)\bmk/k^2$ and $\bmp_B=\bmp-\bmp_E$.
The $E$-mode displacement can be completely described by its divergence 
$\delta_E=-\nabla\cdot\bmp_E$. 
To see how the $E$-mode displacement correlates with the initial linear 
displacement, we carry out a further decomposition,
\bea
\bmp_E(\bmq)=\bmp_E^{(l)}(\bmq)+\bmp_E^{(nl)}(\bmq),
\eea
where
\bea
\bmp_E^{(l)}(\bmk)=W(k)\bmp_L(\bmk),
\eea
and $W(k)={P_{\delta_E\delta_L}(k)}/{P_{\delta_L}(k)}$.
Here, the superscript ``$l$'' denotes the part completely correlated with the 
linear displacement and ``$nl$'' denotes the part generated in nonlinear 
evolution.
Then the nonlinear displacement is composed of three components
\bea
\bmp(\bmq)=\bmp^{(l)}_E(\bmq)+\bmp^{(nl)}_E(\bmq)+\bmp_B(\bmq).
\eea
To measure the power spectra of these three components, we use two sets of simulations. The large box simulation involves $1024^3$ dark matter particles in a 
cubic box of length $600\ h/\mr{Mpc}$ and the small box simulation involves
$1024^3$ dark matter particles in a cubic box of length $150\ h/\mr{Mpc}$.
Each set of simulations has ten independent realizations with independent 
random initial conditions. The power spectrum of displacement is defined as 
\bea
P_\psi(k)=\sum_iP_{\psi_i}(k),
\eea
where $i$ denotes $x$, $y$, or $z$ components.
Figure \ref{fig:pk_disp} shows the power spectra measured from the two sets of
simulations. The displacement power spectra are averaged over ten realizations.
The nonlinear displacement field on large scales is well described by the 
Zeldovich displacement for $k\lesssim1\ h/\mr{Mpc}$.
The nonlinear $E$-mode displacement dominates over the linear $E$-mode displacement at $k=1\ h/\mr{Mpc}$. The $B$-mode displacement is negligible on scales 
larger than $k=10\ h/\mr{Mpc}$.

\section{Lagrangian velocity field}
\label{appendix:B}

\begin{figure}[tbp]
\begin{center}
\includegraphics[width=0.48\textwidth]{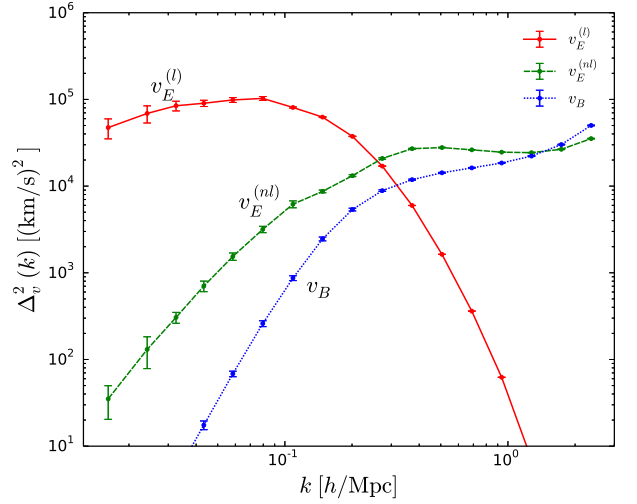}
\end{center}
\vspace{-0.7cm}
\caption{The power spectra for the decomposed Lagrangian velocity. The linear
    part of the velocity is dominant at $k\lesssim0.25\ h/\mr{Mpc}$. 
    The nonlinear contribution becomes important at $k=0.25\ h/\mr{Mpc}$. 
    The curl part is already non-negligible on scales where the displacement 
    can still described by the gradient.}
\label{fig:pk_vel}
\end{figure}

The Lagrangian velocity describes the growth of the Lagrangian displacement,
\bea
\bmv(\bmq)=\dot{\bmp}(\bmq),
\eea
where the dot denotes partial derivative with respect to time.
Since both fields are defined in Lagrangian space, we can decompose the 
Lagrangian velocity similarly,
\bea
\bmv(\bmq)=\bmv_E(\bmq)+\bmv_B(\bmq),
\eea
where $\bmv_E=(\bmk\cdot\bmv)\bmk/k^2$ and $\bmv_B=\bmv-\bmv_E$. 
The $E$-mode displacement can be decomposed into
\bea
\bmv_E(\bmq)=\bmv_E^{(l)}(\bmq)+\bmv_E^{(nl)}(\bmq)
\eea
where 
\bea
\bmv_E^{(l)}(\bmk)=W_v(k)\bmv_L(\bmk).
\eea
Here, $\bmv_L(\bmk)$ is derived from the linear density field,
\bea
\bmv_L(\bmk)=afH\frac{i\bmk\delta_L(\bmk)}{k^2},
\eea
and the velocity window function is 
\bea
W_v(\bmk)=\frac{P_{\theta_E\delta_L}(k)}{P_{\delta_L}(k)},
\eea
where $\theta_E=-\nabla\cdot\bmv/(afH)$. 
Now we can write the Lagrangian velocity as
\bea
\bmv(\bmq)=\bmv_E^{(l)}(\bmq)+\bmv_E^{(nl)}(\bmq)+\bmv_B(\bmq),
\eea
where different velocities describe the growth of the corresponding Lagrangian displacements.
The velocity power spectrum is defined as
\bea
P_v(k)=\sum_iP_{v_i}(k),
\eea
In Fig. \ref{fig:pk_vel}, we show the velocity power spectra for different velocities. We only present the results from the large box simulations mentioned above since the Lagrangian velocity is much more nonlinear than the displacement.
The linear theory can only describe the Lagrangian velocity up to $k\lesssim0.25\ h/\mr{Mpc}$. The nonlinearities dominate on scales where the displacement can 
still be described by linear theory. The $B$-mode Lagrangian velocity is already
non-negligible on scales $k\gtrsim1.4\ h/\mr{Mpc}$. 
This is simply because the Lagrangian velocity corresponds to the most 
nonlinear contribution to the Lagrangian displacement as we discussed above. 

The fact that the Lagrangian velocity is more nonlinear than the displacement
implies that it is more important to study the nonlinearity of the velocity
rather than the displacement to model the reconstructed density field.
This also implies that we can use the much more linear displacement to calculate
the higher order terms of the velocity field using perturbation theories, in 
order to construct $\mathcal{O}(2)$ peculiar velocity reconstruction schemes.

\section{Shift velocity field}
\label{appendix:C}

\begin{figure}[tbp]
\begin{center}
\includegraphics[width=0.48\textwidth]{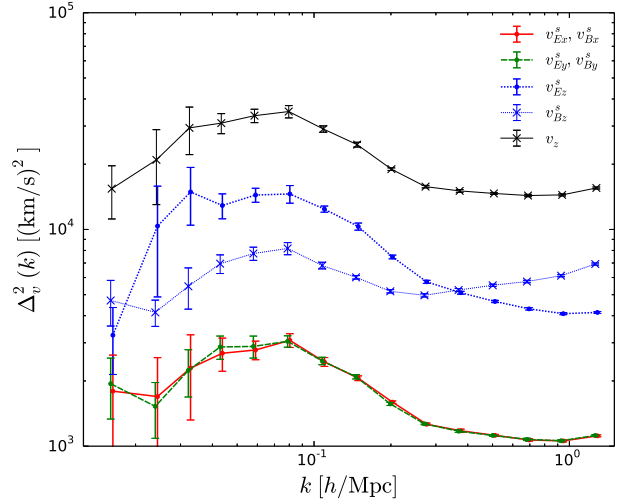}
\end{center}
\vspace{-0.7cm}
\caption{The power spectra for the different shift velocity components.
The power spectrum of ${v^s_{Bz}}$ matches the linear theory prediction at 
$k\lesssim0.1\ h/\mr{Mpc}$, while the other power spectra can be well described 
by linear perturbation theory at $k\lesssim0.15\ h/\mr{Mpc}$. 
The power of $v^s_{Bz}$ becomes larger on smaller scales and dominates over 
$v^s_{Ez}$ at $k=0.36\ h/\mr{Mpc}$.}
\label{fig:pk_vels}
\end{figure}

\begin{figure}[tbp]
\vspace{-0.5cm}
\begin{center}
\includegraphics[width=0.48\textwidth]{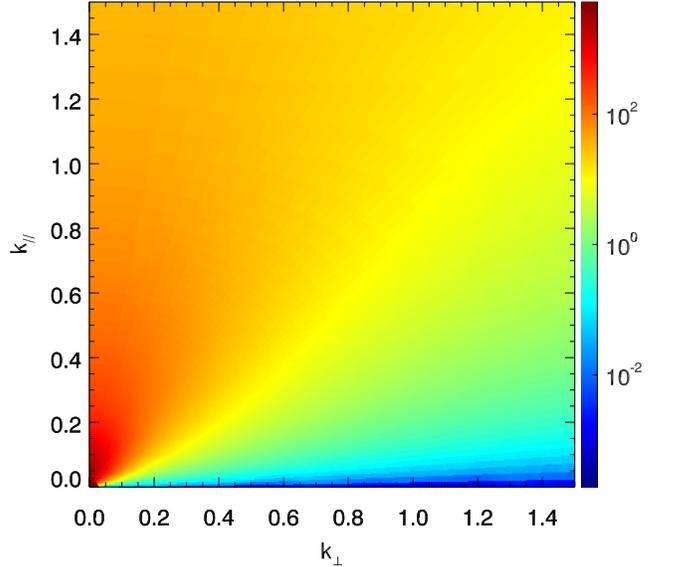}
\end{center}
\vspace{-0.7cm}
\caption{The two-dimensional power spectrum of the shift velocity divergence.
Most power is concentrated in the $\mu$ bins close to $\mu\sim1$ as expected.
The values of the power spectrum are shown in the logarithmic scale.}
\label{fig:pk_deltaV}
\end{figure}

The shift velocity corresponds to the $z$-component of the Lagrangian velocity,
\bea
\bmv^s(\bmq)=v_z(\bmq)\hat{z},
\eea
which describes the shift of a particle from the real space position to the 
redshift space position. We can write the shift velocity as 
\bea
\bmv^s(\bmq)=\bmv_E^s(\bmq)+\bmv^s_B(\bmq),
\eea
where 
\bea
\bmv^s_E(\bmk)=\bigg(\frac{k_xk_zv_z(\bmk)}{k^2},\frac{k_yk_zv_z(\bmk)}{k^2},\frac{k_zk_zv_z(\bmk)}{k^2}\bigg),
\eea
and 
\bea
\bmv^s_B(\bmk)=\bigg(-\frac{k_xk_zv_z(\bmk)}{k^2},-\frac{k_yk_zv_z(\bmk)}{k^2},\frac{(k^2-k_z^2)v_z(\bmk)}{k^2}\bigg).\nonumber\\
\eea
In linear theory, we can compute the power spectra of these velocities from the 
linear initial density.
From the linear continuity equation, we have 
\bea
v_z(\bmk)=\frac{ik_z}{k^2}afH\delta_L(\bmk).
\eea
Then the angular averaged power spectrum of $v_z(\bmk)$ is
\bea
P_{v_z}(k)=\frac{1}{3}(afH)^2P_{\delta_L}(k)/{k^2}.
\eea
In linear theory, the power spectra of different velocity components are
\bea
P_{v^s_{Ex}}(k)=P_{v^s_{Ey}}(k)=\frac{3}{35}P_{v_z}(k),
\eea
\bea
P_{v^s_{Ez}}(k)=\frac{15}{35}P_{v_z}(k),
\eea
\bea
P_{v^s_{Bx}}(k)=P_{v^s_{By}}(k)=\frac{3}{35}P_{v_z}(k),
\eea
and
\bea
P_{v^s_{Bz}}(k)=\frac{8}{35}P_{v_z}(k).
\eea
Figure \ref{fig:pk_vels} shows the velocity power spectra measured from the 
large box simulations. The effects of the $x$- and $y$-components of the $E$-
and $B$-mode shift velocities are much smaller than the $z$-components. 
The power spectrum of ${v^s_{Bz}}$ matches the linear theory prediction at 
$k\lesssim0.1\ h/\mr{Mpc}$, while the other power spectra can be well described 
by linear theory at $k\lesssim0.15\ h/\mr{Mpc}$. The power of $v^s_{Bz}$ becomes
larger on smaller scales and dominates over $v^s_{Ez}$ at $k=0.36\ h/\mr{Mpc}$.
This behavior is not predicted within the linear perturbation theory. 
The strong nonlinearities in the shift velocity field cause the large power of 
$P_{v^s_{Bz}}(k)$ on small scales. This implies that the random velocity dispersion is large on 
small scales, which leads to the so-called finger of God effect.

Figure \ref{fig:pk_deltaV} shows the two-dimensional power spectrum of the 
negative divergence of the shift velocity $-\nabla\cdot\bmv^s$.
We plot the power spectrum in logarithmic scale for visualization purposes.
Most power of $-\nabla\cdot\bmv^s$ is concentrated in the $\mu$ bins close to 
$\mu\sim1$ as expected. This causes part of the degradation of reconstruction
performance along the line of sight direction.

\section{Reconstruction with halos}
\label{appendix:D}

\begin{figure*}[tbp]
\vspace{-0.3cm}
\begin{center}
\includegraphics[width=0.32\textwidth]{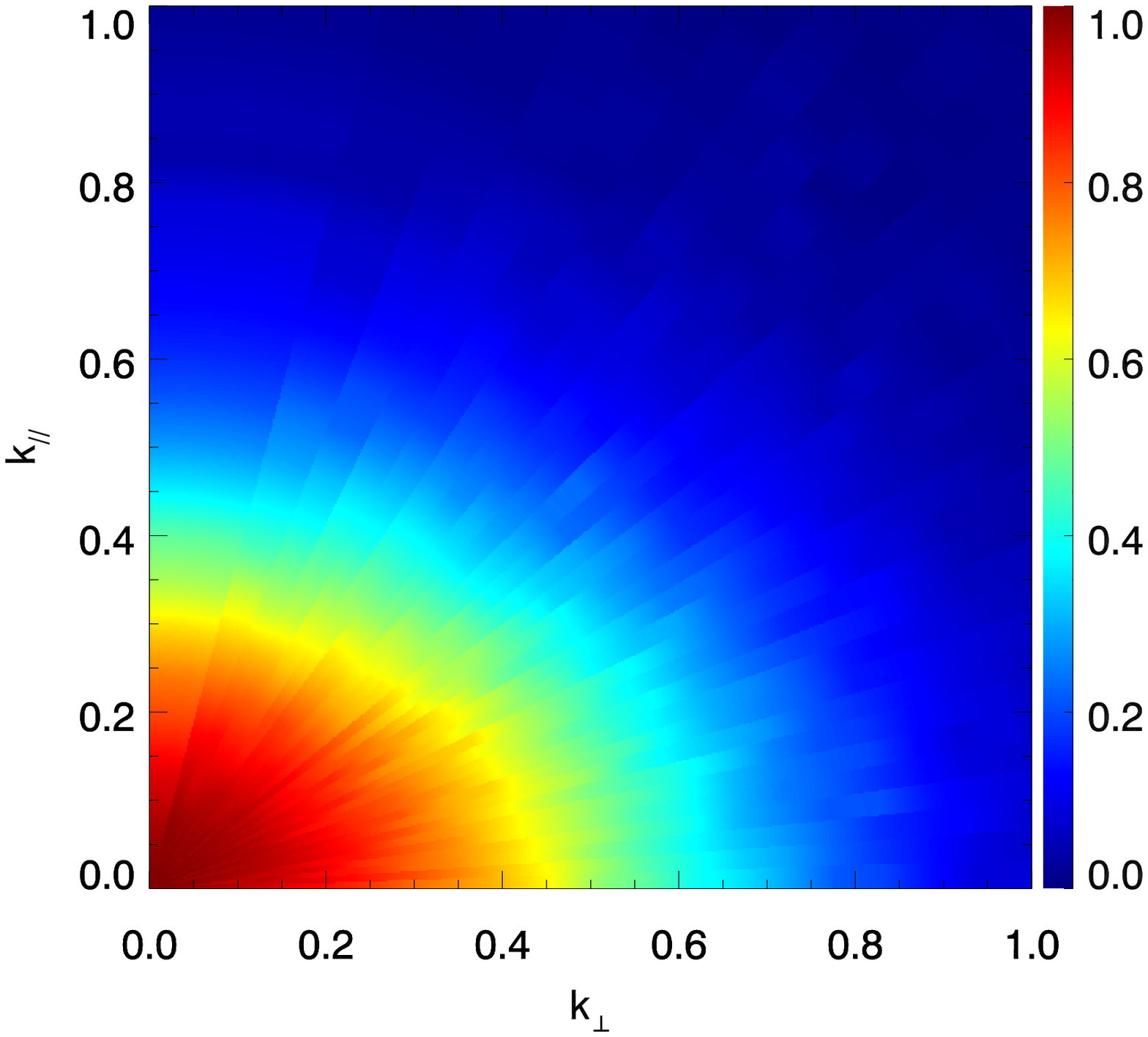}
\includegraphics[width=0.32\textwidth]{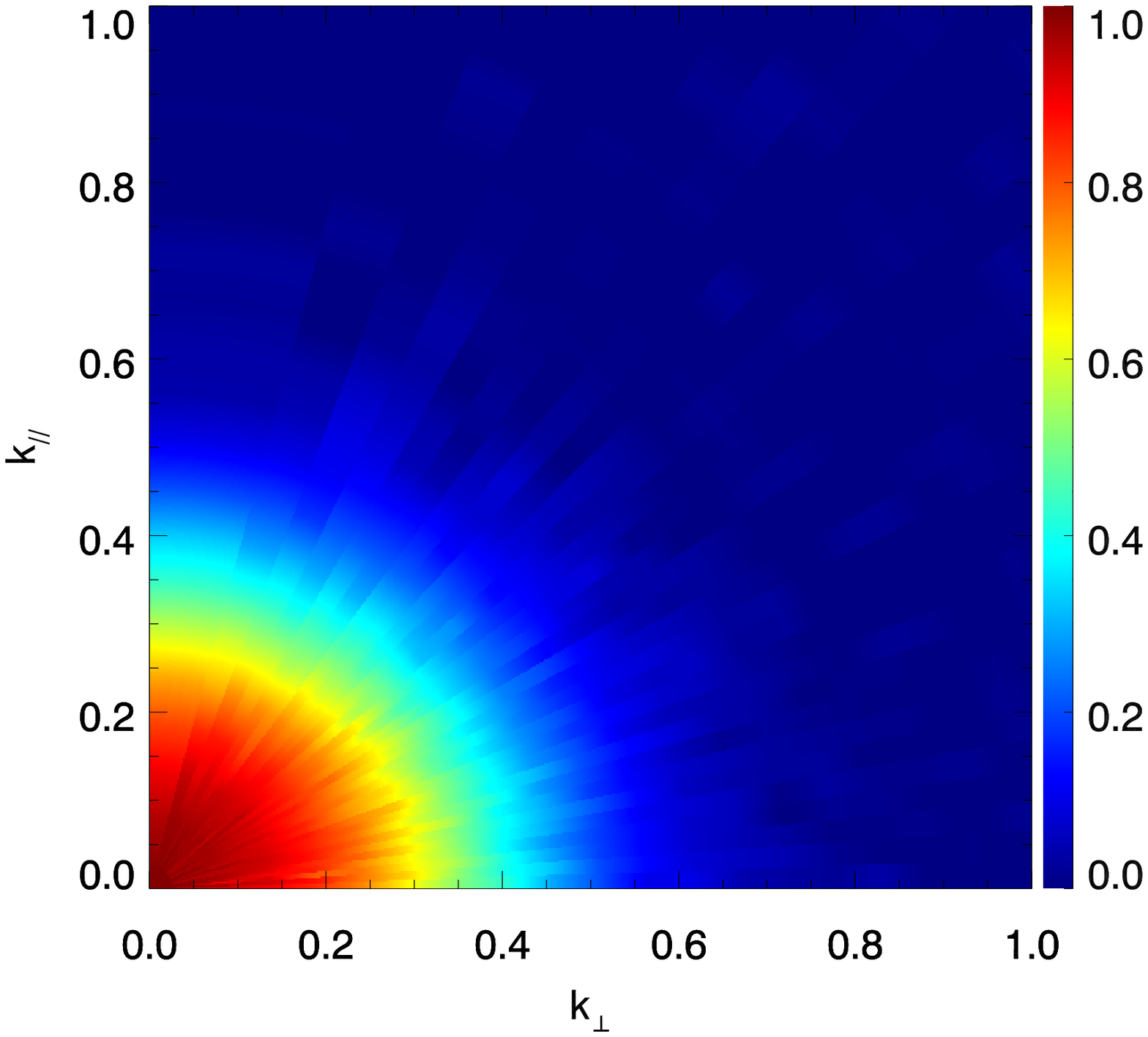}
\includegraphics[width=0.32\textwidth]{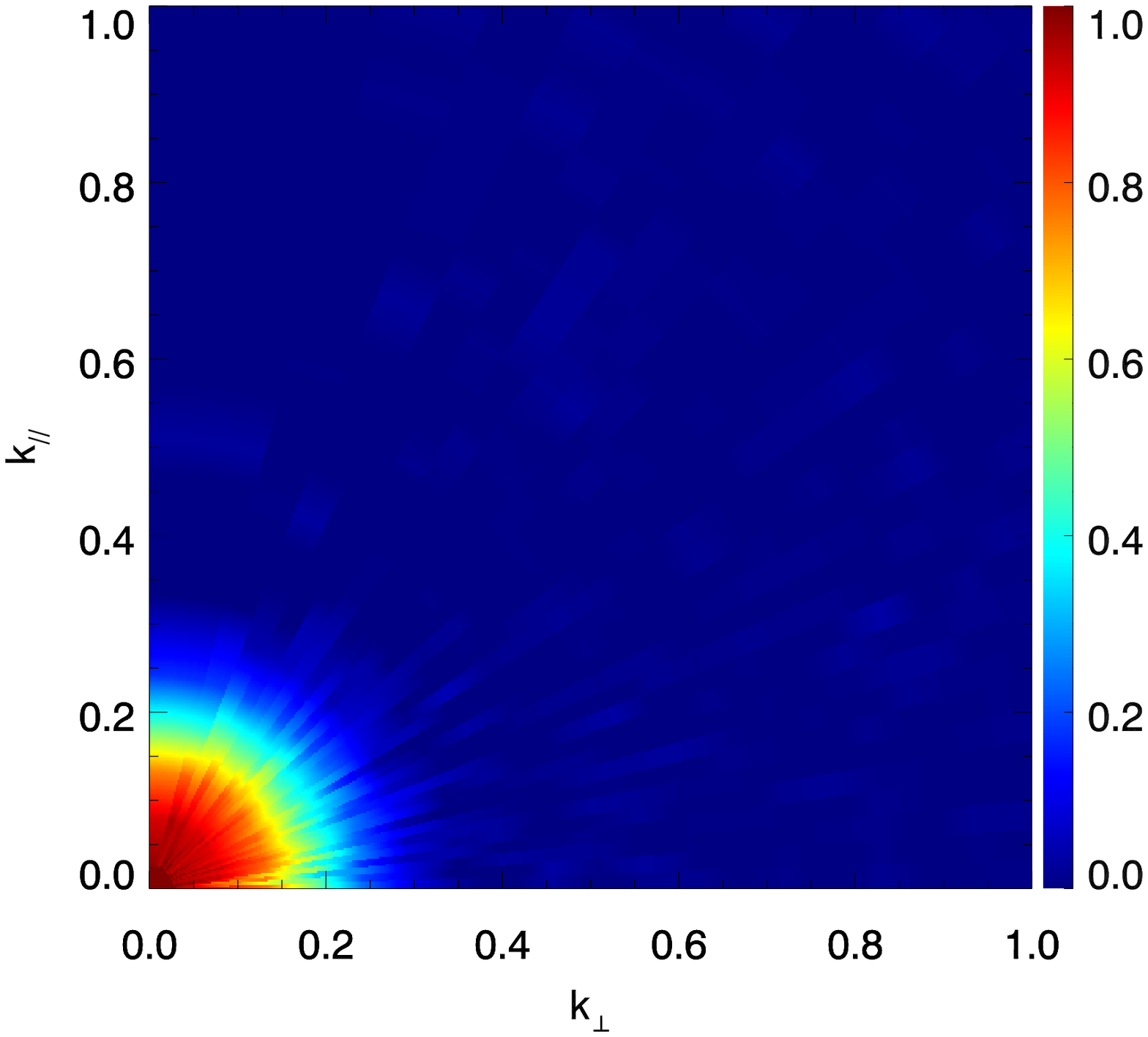}
\end{center}
\vspace{-0.7cm}
\caption{The two-dimensional cross-correlation coefficients of the halo 
reconstructed field with the linear field $(b_1+f\mu^2)\delta_L(\bmk)$ for
$\bar{n}=2.77\times10^{-2}$, $2.77\times10^{-3}$, $2.77\times10^{-4}\ (h/\mr{Mpc})^3$ (from left to right), respectively.}
\label{fig:xcc_ha}
\end{figure*}

The dark matter field is nearly constant in Lagrangian space, while the halo 
field is not uniform in Lagrangian space. 
However, when we apply reconstruction,
we still map the halo density field into a nearly uniform distribution as
\bea
\rho_h(\bmx)d^3x=\rho(\bmq)d^3q,
\eea
where $\rho_h(\bmx)$ is the halo density field in Eulerian space. As a result,
the reconstructed displacement is biased,
\bea
\bmp_h(\bmk)=b(k)\bmp(\bmk),
\eea
where $b(k)$ is the bias factor of the displacement reconstructed from the 
halo field. At linear order the displacement bias is equal to the constant 
Eulerian halo bias,
\bea
\delta_h(\bmx)=b_1\delta(\bmx),
\eea
where $b_1$ is the linear halo bias. However, the mapping from real to redshift
space for the halo density field is still defined as the dark matter field
\bea
\rho_h(\bms)d^3s=\rho_h(\bmx)d^3x,
\eea
where $\rho_h(\bms)$ is the redshift space halo density field. 
Thus the displacement reconstructed from halos in redshift space is 
\bea
\bmp^s_h(\bmk)=b(k)\bmp(\bmk)+{\bmv^s(\bmk)}/{aH}.
\eea
To quantify the performance of reconstruction with halos in redshift space,
we should correlate the reconstructed field with $(b_1+f\mu^2)\delta_L(\bmk)$
instead of $(1+f\mu^2)\delta_L(\bmk)$.

Figure \ref{fig:xcc_ha} shows the two-dimensional cross-correlation coefficients
of the halo reconstructed density field with the linear field $(b_1+f\mu^2)\delta_L(\bmk)$.
Three halo samples with number densities of $2.77\times10^{-2}$, $2.77\times10^{-3}$, and $2.77\times10^{-4}\ (h/\mr{Mpc})^3$ are used for reconstruction.
The bias factors are $b_1=0.68$, $0.92$, and $1.44$, respectively.
We find that the correlation coefficients have the same anisotropic features 
as the result obtained with the dark matter density field, but the performance 
degrades with the lower number density. And also the less dense sample suffers
less from the RSD effect.

\section{Reconstruction with survey mask}
\label{appendix:E}

\begin{figure}[tbp]
\vspace{-0.7cm}
\begin{center}
\includegraphics[width=0.48\textwidth]{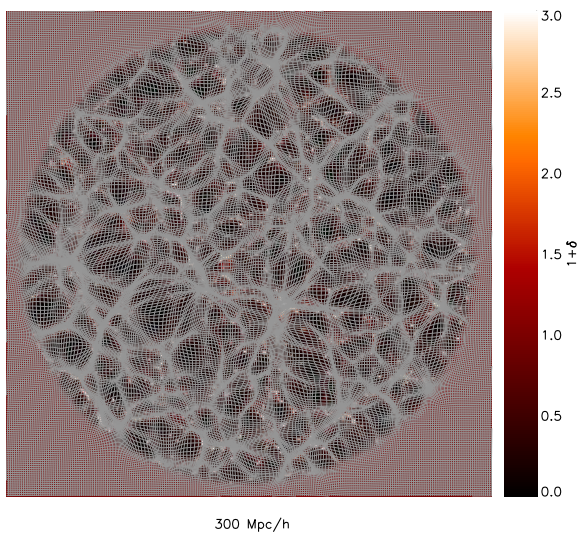}
\end{center}
\vspace{-0.7cm}
\caption{A slice of the density field from the high-resolution simulation. 
The curvilinear grid shows the Eulerian coordinate of each grid point of the 
potential isobaric coordinate. The reconstruction works very well within the 
spherical region though near the spherical boundary there are some artificial 
features caused by the discontinuities.}
\label{fig:mask}
\end{figure}

To apply nonlinear reconstruction to real observation data, we need to consider
the boundary of a survey volume. In this appendix  we present a simple test of 
reconstruction without the periodic boundary conditions. We take a spherical 
region with radius $140.625\ \mr{Mpc}/h$ from the high-precision simulation and
set the density fluctuations $\rho(\bmx)/\bar{\rho}$ in the outer space to 1.
Then we perform reconstruction with this density field.
Figure \ref{fig:mask} shows a slice through the density field. We also plot the 
Eulerian positions of a uniform grid of potential isobaric coordinates on the
density field slice. The reconstruction works very well within the spherical
region though near the spherical boundary there are some artificial features
caused by the discontinuities. However, this causes little degradation of 
reconstruction as the measure of a two-dimensional boundary is nearly $0$ 
in the three-dimensional space.

\bibliographystyle{apsrev}
\bibliography{3d}

\end{document}